\documentclass[11pt]{article}
\usepackage{graphicx}
\usepackage{amsmath}

\input{amssym.def}
\input{amssym.tex}

\usepackage{amscd}

\allowdisplaybreaks
 \numberwithin{equation}{section}

\begin{document}

\begin{titlepage}
\thispagestyle{empty}
\begin{flushleft}
\hfill hep-th/0306189 \\
UT-03-22\hfill June, 2003 \\
\end{flushleft}

\vskip 1.5 cm
\bigskip

\begin{center}
\noindent{\LARGE Boundary states as exact solutions}\\
\noindent{\LARGE of (vacuum) closed string field theory}\\

\renewcommand{\thefootnote}{\fnsymbol{footnote}}

\vskip 2cm
{\large
Isao~Kishimoto\footnote{e-mail address:
 ikishimo@hep-th.phys.s.u-tokyo.ac.jp}, 
Yutaka~Matsuo\footnote{e-mail address:
 matsuo@phys.s.u-tokyo.ac.jp} and  
Eitoku~Watanabe\footnote{e-mail address:
 eytoku@hep-th.phys.s.u-tokyo.ac.jp}}
\\
{\it
\noindent{ \bigskip }\\
Department of Physics, Faculty of Science, University of Tokyo \\
Hongo 7-3-1, Bunkyo-ku, Tokyo 113-0033, Japan\\
\noindent{ \smallskip }\\
}
\vskip 20mm
\bigskip
\end{center}
\begin{abstract}
We show that the boundary states are idempotent
$B*B = B$ with respect to the star product of
HIKKO type  {\em closed} string field theory. Variations around
the boundary state  correctly reproduce the open string spectrum
with the gauge symmetry. We explicitly demonstrate it for
the tachyonic and massless vector modes.
The idempotency relation may be regarded as the
equation of motion of
closed string field theory at a possible vacuum.
\end{abstract}
\end{titlepage}\vfill\setcounter{footnote}{0} 
\renewcommand{\thefootnote}{\arabic{footnote}} 

\newpage


\section{Introduction
\label{sec:Intro}
}

Study of the off-shell structure of
string theory is  an essential step 
in understanding
its non-perturbative physics.
In recent years, Witten-type open string field theory \cite{Witten}
has been intensively examined in this context.
One of the goals is to understand D-branes
as soliton solutions of open string field theory.
One of the promising discoveries was that the energy of
the tachyon vacuum correctly reproduced the tension
of D-branes at least numerically.\cite{G-R}

Inspired by the experiences of noncommutative field theory,
it was conjectured by Rastelli, Sen and Zwiebach that
the D-branes may be understood as the solutions to
the projector equation,
\begin{equation}\label{eq_projector}
 \Psi\star\Psi = \Psi\,\,,
\end{equation}
where $\star$ is the noncommutative and associative
Witten-type star product for 
an open string field.
It was conjectured that this equation may be understood
as the equation of motion of 
a string field expanded around
the tachyon vacuum (the 
so-called 
vacuum string field theory (VSFT) 
conjecture \cite{VSFT}\cite{GRSZ}).
In particular, a few examples of the projectors, the sliver
state or 
butterfly state, were examined as the candidates
which describe the D-brane.

It turned out, however, that the treatment of D-branes
in open string field theory is very delicate.
One of the difficulties was the description of the closed string sector.
In Witten-type open string field theory, the action does not include the
closed string degrees of freedom at the tree level.
If we need to describe them in open string language alone, we have
to consider a singular state such as identity string field where the 
closed string vertex is inserted at the
midpoint.\cite{Strominger:1986zd}\cite{Strominger:ad}\cite{GRSZ} 
The midpoint in open string field theory causes
many subtleties, for example, it causes the breakdown of
the associativity \cite{Horowitz:yz}
 and we have to be very careful 
while handling
such a degree of freedom.\footnote{
Recently, a regularization method 
was proposed \cite{BM1}.
}
D-brane couples to the closed string sector 
(for example, gravity) at the tree level, and we cannot escape 
from using such a  singular description.  
The level truncation regularization
seems to handle it numerically.
However, the analytic treatment of the problem remains 
as a real challenge.

In this paper, we change the viewpoint
and start the analysis of D-branes
{}in {\em closed string field theory}.
We believe that such a treatment is natural since
the nature of D-branes is most precisely encoded in the
boundary state $|B\rangle$ which lives in the Hilbert
space of the closed string sector.
In particular, we will prove that the boundary states 
(both for Neumann and Dirichlet boundary conditions)
satisfy an analogue of Eq.(\ref{eq_projector}),
\begin{equation}\label{eq_projector2}
 |B\rangle * |B\rangle = |B\rangle\,\,
\end{equation}
up to a pure ghost prefactor.

\begin{figure}[htbp]
	\begin{center}
	\scalebox{0.5}[0.5]{\includegraphics{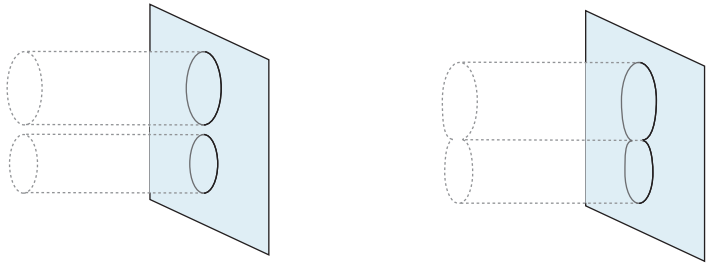}}
	\end{center}
	\caption{$*$ product of the boundary states}
	\label{fig:bs}
\end{figure}

Unlike the open string version, 
Eq.(\ref{eq_projector2}) has a natural geometrical meaning.
The boundary state, as suggested by its name, describes the
boundary condition of the string world sheet.
Suppose there exist two holes with the same type of boundary
condition.  If we merge these two holes by 
a closed string
star product, we expect to have the same boundary condition
on the new hole.(Fig.\ref{fig:bs})

To demonstrate this observation explicitly, we have to be
specific about the choice of the star product.
There are three candidates of closed string field theory
which were well examined so far.

The oldest one is the light-cone gauge approach \cite{ref:LCG}.
This is consistent in the sense that it produces the correct
integration range over the moduli parameter.  
However, for the application to our problem, it is not useful
since the boundary states have nontrivial dependence on the 
time coordinate.  We need covariant descriptions.

The second one is the closed string version of
Witten's open string theory.
A generalization of Witten-type 
midpoint interaction vertex to closed strings results in
nonpolynomial string field theory \cite{S-Z,K-S}\footnote{See
 \cite{Zwiebach} for a review.}.
The action contains infinitely many terms to cover the moduli spaces
for the Riemann surfaces corresponding to various interactions. 
This approach contains many mathematically interesting
features such as $L_\infty$ structure.
Handling of the moduli parameters still remains
as a challenge, however, and it has not reached the 
completely satisfactory level.

The third one is based on a split-joining type vertex,
which was proposed about the same time
as Witten-type open string field theory
and is now known as HIKKO's 
(Hata-Itoh-Kugo-Kunitomo-Ogawa)
string field theory \cite{HIKKO1,HIKKO2}.
It has exactly the same action as Witten's open string field theory,
namely, the kinetic term and a three string interaction.\footnote{
 The action for open strings contains 3-string and 4-string vertices
 besides a kinetic term.
}
In HIKKO's theory, it is necessary 
to introduce a parameter called
string length $\alpha$ to specify string interactions, 
which has no analogue
in Witten-type string field theories.
It must be integrated
in computing physical quantities and might cause
a divergence in loop amplitudes \cite{HIKKO:loop}.
The simplest way to resolve this difficulty is to just set $\alpha =
p^+$, but it breaks the covariance.

To summarize, there is no completely satisfactory 
closed string field theory.
In this paper, we adopt HIKKO's star product to explicitly demonstrate
Eq.(\ref{eq_projector2}). 
However, we expect it to hold even if we replace it 
with a Witten-type product.
We will come back to
prove it in our future paper \cite{KMW2}.
We would like to propose this relation as a universal 
characterization of the boundary states in closed string
field theory, which is independent of the specific
proposals for the action. A merit to use HIKKO's
approach is the analogy of
the action with Witten's open string field theory.
If we want to have an analogy with VSFT proposal, 
this gives a good reason to start from it.



We note that HIKKO's $*$ product in Eq.(\ref{eq_projector2}) has 
different properties compared with Witten's star product
in open string field theory.  It may be summarized as the
following relations:
\begin{eqnarray}
&& \Phi * \Psi = -(-1)^{|\Phi||\Psi|} \Psi * \Phi\,\,,
\label{eq_commutativity}
\\
&& (\Phi * \Psi)*\Lambda + 
(-1)^{|\Phi|(|\Psi|+|\Lambda|)}(\Psi * \Lambda) * \Phi
+(-1)^{|\Lambda|(|\Phi|+|\Psi|)}(\Lambda * \Phi) * \Psi=0\,\,,
\label{eq_Jacobi}
\\
&& Q(\Phi * \Psi) = Q\Phi * \Psi +(-1)^{|\Phi|}\Phi * Q\Psi\,\,.
\label{eq_derivative}
\end{eqnarray}
First of all,  the product is (anti-)commutative
(\ref{eq_commutativity}).
While it breaks associativity, it satisfies the analogue of
Jacobi identity (\ref{eq_Jacobi}). In a sense, it has the same
property as the commutator of Witten-type open string product,
$
 \Phi *^{HIKKO} \Psi \leftrightarrow \Phi\star^{Witten}\Psi 
-(-1)^{|\Phi||\Psi|}
 \Psi \star^{Witten} \Phi\,\,.
$
Since the nature of the product is different, we cannot
interpret the equation (\ref{eq_projector2}) as
defining a projector.  
In the following, however, we will continue to use
the word ``projector'' to describe the state that satisfies
Eq.(\ref{eq_projector2}) because of the similarity with the
discussion of the open string.

We conjecture that Eq.(\ref{eq_projector2}) gives a good characterization
of the conformal invariant boundary. For this purpose, we
calculate an infinitesimal variation of the boundary state
of the following form,
\begin{equation}\label{eq_insertion}
 \delta_V |B\rangle = \oint d\sigma 
 V(\sigma) |B\rangle
\end{equation}
where $V(\sigma)$ is a vertex operator inserted at the boundary.
We argue that the idempotency condition (\ref{eq_projector2})
requires the vertex $V$ to be marginal.
We will prove this expectation for 
the tachyonic state and the massless vector
state. For such  variations, this gives the
mass-shell condition for these open string modes.
In a sense, the idempotency condition knows the mass-shell condition
of the open string while they are the equation for the closed string
states!

We note that our argument is very similar to the discussion
of vacuum string field theory.  For example, use of
the variation of Eq.(\ref{eq_projector2}) to derive
the mass-shell condition for the open string states was
examined in the VSFT context by Hata-Kawano \cite{HK}
and Okawa \cite{Okawa}.  In particular, in the latter approach,
the marginal deformation was made 
over the whole boundary.
This is basically the same variation as Eq.(\ref{eq_insertion}).
The difference is, of course, the Hilbert space where the
projector lives.  In VSFT, to describe such an projector, we have to
consider singular states.  For example, the sliver state is
made by taking the infinite star products of the vacuum state.
On the other hand, our closed string description does not include
such a singular manipulation.  The boundary state is a well-defined
state in the boundary conformal field theory.  In this way, we
can escape from the subtleties of VSFT.

The paper is organized as follows.
In section \ref{sec:claims}, we give the explicit definitions of
the boundary states and the 3-string vertex  
which are discussed in this paper.  We will then present
our  claims more precisely. The proof is given
explicitly in the following sections which are rather technical.
In section \ref{sec:Gaussian} we prove the idempotency relation
of the boundary states.  We need many properties of the
Neumann coefficients which are summarized in appendix \ref{sec:formulae}.
In section \ref{sec:fluctuation} we investigate infinitesimal variations
around the boundary state and derive on-shell condition of open string
on them. 
In section \ref{sec:discussion}, we discuss some issues of our results.

\section{Boundary state and star product of closed string field theory}
\label{sec:claims}
\subsection{Boundary states}
The boundary states $|B(F)\rangle$ which we are going to discuss
are those for D$p$-branes with constant field strength $F_{\mu\nu}$
\cite{Callan},
\begin{eqnarray}
 |B(F)\rangle&=&e^{-\sum_{n\ge 1}a^{(+)\dagger}_n{\cal O}a^{(-)\dagger}_n}
e^{\sum_{n\ge 1}(c^{(+)\dagger}_n\bar{c}^{(-)\dagger}_n
+c^{(-)\dagger}_n\bar{c}^{(+)\dagger}_n)}
|p_{\mu}=0,x^i\rangle\,,
\label{eq:B(F)}
\\
 &&{\cal O}^{\mu}_{~\nu}=\left[(1+F)^{-1}(1-F)\right]^{\mu}_{~\nu}\,,
~~~\mu,\nu=0,1,\cdots, p\,,\\
 &&{\cal O}^{i}_{~j}=-\delta^i_j\,,~~~~i,j=p+1,\cdots, d-1 \,\,\,.
\end{eqnarray}
$x^{\mu}\,(\mu=0,1,\cdots,p)$ are the coordinates
along the Neumann directions and 
$x^i\,(i=p+1,\cdots, d-1)$ are along the Dirichlet directions\footnote{
We summarize our notation of the oscillators and the 
vacuum state in appendix  \ref{sec:conventions}.
In particular, we use $\bar{c}$ to denote the anti-ghost
(usually written as $b$) by following HIKKO's convention.
\cite{HIKKO1,HIKKO2}
For ghost zero mode convention,
we use  $\pi_c^0$-omitted formulation
(section VB.in Ref. \cite{HIKKO2}).
}.
We use the letters $M,N$ ($=0,\cdots,d-1$)
to represent all these directions.
We put $d=26$ since we are considering bosonic string theory.
These states  satisfy the following 
conditions:
\begin{eqnarray}
&&\sqrt{\pi}X^i(\sigma) |B(F)\rangle=x^i|B(F)\rangle\,,\label{eq:BFi}\\
&&\left(P_{\mu}(\sigma)-F_{\mu\nu}{d\over d\sigma}X^{\nu}(\sigma)\right)
|B(F)\rangle=0\,,\label{eq:BFmu}\\
&&\pi_c(\sigma)|B(F)\rangle=\pi_{\bar{c}}(\sigma) |B(F)\rangle=0\,.
\label{eq:BFpi}
\end{eqnarray}
The boundary states are invariant under BRST transformation
 $Q_B|B(F)\rangle=0$.\footnote{This property is essential
to couple the boundary state as the external source  to
closed string field theory.
The authors of Ref. \cite{HH} proposed such an action
\begin{equation}
S_{\rm tot}= {1\over g^2}\left\{{1\over 2}\Phi\cdot Q_B\Phi
+{1\over 3}\Phi\cdot(\Phi*\Phi)\right\}+B(F)\cdot \Phi+I(F)\,,
\end{equation}
namely, $Q_B|B(F)\rangle=0$ is necessary to satisfy the 
gauge invariance of $S_{\rm tot}$.
This was the first example where the boundary state appeared
essentially in closed string field theory.
They used this action to derive open string action 
(Born-Infeld action) and proved their gauge invariance
through string field theory.
An unsatisfactory point was, however, that one needs to put
the boundary state by hand from outside.
Our study starts from a hope to {\em derive} it
within the framework of closed string field theory.
}
${\cal O}$ is  orthogonal $\mathcal{O}\mathcal{O}^T=
\mathcal{O}^T\mathcal{O}=1$ 
since $F_{\mu \nu}$ is antisymmetric.
Along the Dirichlet directions, this matrix
becomes trivial in the sense: $\left({1+{\cal O}\over 2}\right)^i_{~j}=0$,
but zero modes have nonzero momentum.
The oscillator representations of Eqs.(\ref{eq:BFi}--\ref{eq:BFpi})
are summarized in the appendix \ref{sec:Oscil Boundary}.


\subsection{Reflector and 3-string vertex}
HIKKO's star product for the closed string is a covariant
version of light-cone string field theory.
It is defined by the reflector $\langle \tilde{R}|$ which maps
a ket vector to a bra vector 
and the three string vertex $|V(1,2,3)\rangle$
which lives in
the tensor product of three closed string Hilbert spaces :
\begin{eqnarray}
\label{eq:star-dfn}
 |\Phi_1 * \Phi_2\rangle_3&=&\int d\bar{c}^{(1)}_0 d\bar{c}^{(2)}_0
{}_1\langle\Phi_1|{}_2\langle\Phi_2 |V(1,2,3)\rangle ,\\
 {}_2\langle \Phi|&:=&\int d\bar{c}_0^{(1)}
\langle \tilde{R}(1,2)|\Phi\rangle_1\,\,.
\end{eqnarray}
The reflector is defined by \cite{HIKKO2}:
\begin{eqnarray}
\langle \tilde{R}(1,2)|&=&\int d^dx^{(1)}d^dx^{(2)}
{d\alpha_1 d\alpha_2\over(2\pi)^2} {}_1\langle x^{(1)},\alpha_1|
{}_2\langle x^{(2)},\alpha_2|
\nonumber\\
&&\times \exp\left[-\sum_{\pm,n\ge 1}\left(
a_n^{(\pm)(1)}\cdot a_n^{(\pm)(2)}
+c_n^{(\pm)(1)}\bar{c}_n^{(\pm)(2)}
-\bar{c}_n^{(\pm)(1)}c_n^{(\pm)(2)}\right)\right]
\nonumber\\
&&\times \delta^d(x^{(1)}-x^{(2)})\delta(\bar{c}_0^{(1)}-\bar{c}_0^{(2)})
2\pi\delta(\alpha_1+\alpha_2)\,.
\end{eqnarray}
The 3-string vertex $|V(1,2,3)\rangle$ is given explicitly in terms of
oscillators  as\footnote{
This is the same as $|V'(1,2,3)\rangle$ in Eq.(5.15) in \cite{HIKKO2},
which is the 3-string vertex in $\pi_c^0$-omitted formulation.
}
\begin{eqnarray}
&&|V(1,2,3)\rangle=\int\delta(1,2,3)[\mu(1,2,3)]^2\wp^{(1)}\wp^{(2)}\wp^{(3)}
\nonumber\\
&&~~~~~~~~~~~~~~~~~~\times
\prod_{r=1}^3\left(1+{1\over \sqrt{2}}w^{(r)}_I\bar{c}_0^{(r)}\right)
e^{F(1,2,3)}|p_1,\alpha_1\rangle_1|p_2,\alpha_2\rangle_2
|p_3,\alpha_3\rangle_3\,,~~~~~
\label{eq:v123}
\\
&&F(1,2,3)=\sum_{\pm}\sum_{r,s=1}^3\sum_{m,n\ge 1}\tilde{N}^{rs}_{mn}
\left({1\over 2}\,a_m^{(\pm)(r)\dagger}\cdot a_n^{(\pm)(s)\dagger}
+\sqrt{m}\alpha_rc^{(\pm)(r)\dagger}_{m}(\sqrt{n}\alpha_s)^{-1}
\bar{c}^{(\pm)(s)\dagger}_{n}\right)\nonumber\\
&&~~~~~~~~~~~~~~~~
+{1\over 2}\sum_{\pm}\sum_{r=1}^3\sum_{n\ge 1}\tilde{N}^r_n \,
a^{(\pm)(r)\dagger}_n \cdot {\bf P}
-{\tau_0\over 4\alpha_1\alpha_2\alpha_3}{\bf P}^2\,,\\
&&{\bf P}:=\alpha_1p_2-\alpha_2p_1\,,\\
&&w_I^{(r)}={1\over \sqrt{2}}\sum_{\pm}\sum_{s=1}^3\sum_{m\ge 1}w^{rs}_m
\alpha_s c^{(\pm)(s)\dagger}_m\,,~~~~
w^{rs}_m=\chi^{rs}m\bar{N}^s_m+{1\over \alpha_r}\sum_{n=1}^{m-1}m
\bar{N}^{ss}_{m-n,n}\,,\\
&&\chi^{rs}=\delta_{r,s}{1\over \alpha_r}(\alpha_{r-1}-\alpha_{r+1})
+\sum_{t=1}^3\epsilon^{rst}\,,~~~~(\alpha_4:=\alpha_1\,,~~\epsilon^{123}=+1)
\\
&&\mu(1,2,3)=\exp\left(-\tau_0\sum_{r=1}^3{1\over \alpha_r}\right)\,,
~~~~~~\tau_0=\sum_{r=1}^3\alpha_r\log |\alpha_r|\,,\\
&&\int\delta(1,2,3)=\int {d^dp_1\over (2\pi)^d}{d^dp_2\over (2\pi)^d}
{d^dp_3\over (2\pi)^d}
{d\alpha_1\over 2\pi}{d\alpha_2\over 2\pi}{d\alpha_3\over 2\pi}
(2\pi)^d\delta^d(p_1+p_2+p_3)2\pi\delta(\alpha_1+\alpha_2+\alpha_3)\,,~~~\\
&&\wp^{(i)}=\oint {d\theta\over 2\pi} \,e^{-i\theta(N_+^{(i)}-N_-^{(i)})}\,,
~~~~
N_{\pm}=\sum_{n\ge 1}n\left(a_n^{(\pm)\dagger}\cdot a_n^{(\pm)}
+c_{n}^{(\pm)\dagger}\bar{c}_n^{(\pm)}+\bar{c}_{n}^{(\pm)\dagger}
c_n^{(\pm)}\right)\,.
\end{eqnarray}
The coefficients $\tilde N^{rs}_{mn}$ are Neumann coefficients; 
$\tilde{N}_{mn}^{rs}:=\sqrt{m}\bar{N}_{mn}^{rs}\sqrt{n}\,,~
\tilde{N}^r_m:=\sqrt{m}\bar{N}^r_m$.
Their definitions and some formulae which they satisfy are
given in appendix \ref{sec:formulae}.
$\wp^{(i)}$ is a projector to impose the level matching condition $N_+=N_-$ on
the $i$th string. 
Note that we can rewrite some of the above as
\begin{equation}
{\bf P}=\alpha_1p_2-\alpha_2p_1=\alpha_2p_3-\alpha_3p_2
=\alpha_3p_1-\alpha_1p_3\,,~~~
-{{\bf P}^2\over 4\alpha_1\alpha_2\alpha_3}
= \sum_{r=1}^3{p_r^2\over 4\alpha_r}
\end{equation}
in the presence of $\delta$-functions which impose
$p_1+p_2+p_3=0$ and $\alpha_1+\alpha_2+\alpha_3=0$.

\begin{figure}[htbp]
	\begin{center}
	\scalebox{0.4}[0.4]{\includegraphics{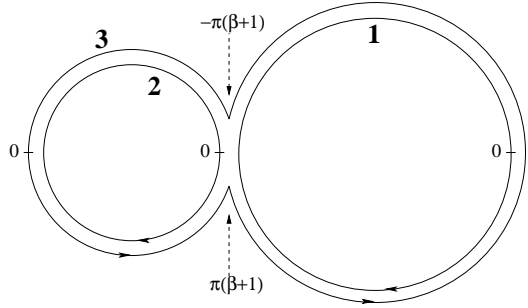}}
	\end{center}
	\caption{Overlapping configuration of three closed strings.
Strings with the labels 1,2,3 whose length parameters are 
$|\alpha_1|,|\alpha_2|,|\alpha_3|(=|\alpha_1|+|\alpha_2|)$
 are parametrized by
 $\sigma_1,\sigma_2,\sigma_3\,(-\pi\le \sigma_r\le \pi)$. 
$\sigma_3=\pm \pi(\beta+1)~~(\beta:=\alpha_1/\alpha_3)$ 
are interaction points on string 3 which correspond to 
$\sigma_1=\pm \pi,\sigma_2=0$ on  string 1 and 2, respectively.
}
	\label{fig:int}
\end{figure}

The 3-string vertex $|V(1,2,3)\rangle$ is determined by
the overlap conditions (Fig.\ref{fig:int}),
\begin{eqnarray}
 && \Theta_1 X^{(1)}(\sigma_1) +\Theta_2 X^{(2)}(\sigma_2)
-X^{(3)}(\sigma_3)=0\,\,,\label{eq_over1}\\
 && \Theta_1 \alpha_1 c^{(1)}(\sigma_1) +\Theta_2 
\alpha_2 c^{(2)}(\sigma_2)
-\alpha_3 c^{(3)}(\sigma_3)=0\,\,,\label{eq_over2}\\
 && \Theta_1 \alpha_1^{-2}\bar{c}^{(1)}(\sigma_1) +\Theta_2 
\alpha_2^{-2} \bar{c}^{(2)}(\sigma_2)
-\alpha_3^{-2} \bar{c}^{(3)}(\sigma_3)=0\,\,,\label{eq_over3}\\
&& \Theta_1(\sigma) \equiv \theta(\pi|\alpha_1|-|\sigma|)\,,
\Theta_2(\sigma) \equiv \theta(|\sigma|-\pi|\alpha_1|)\,,\label{eq_over4}\\
&& \sigma_1(\sigma)=\frac{\sigma}{\alpha_1}\,,\quad
\sigma_2(\sigma) = \frac{\sigma-\mathrm{sgn}(\sigma)\pi|\alpha_1|}{
\alpha_2}\,,\quad 
\sigma_3(\sigma) = \frac{ \mathrm{sgn}(\sigma)\pi|\alpha_3|-\sigma}{
-\alpha_3}\,, \label{eq_over5}
\end{eqnarray}
where $-\pi|\alpha_3|\le \sigma\le \pi|\alpha_3|$  and
$\alpha_i$ ($i=1,2,3$) are real parameters with the constraint
$\alpha_1+\alpha_2+\alpha_3=0$.
In light-cone string field theory, they are interpreted
as the light-cone momenta which are preserved at the interaction.
In the covariant theory, they become the external parameters
which characterize the overlap conditions 
(\ref{eq_over1}--\ref{eq_over3}).
We note that the above conditions are for the particular
case $|\alpha_3|=|\alpha_1|+|\alpha_2|$
and we need some modifications for other choices.

\subsection{Main results}
At this point, it is possible to make a precise
statement of our results, given as follows.
\begin{enumerate}
 \item We slightly redefine the boundary state $|B(F)\rangle$ 
(i) by multiplying $\bar{c}_0$
to obtain the correct ghost number of closed string field
in the physical sector in gauge-fixed action \cite{HIKKO2}
and (ii) by  including the 
string-length parameter ($\alpha$ parameter)
\begin{eqnarray}
 |\Phi_B(\alpha)\rangle&=&
e^{-a^{(+)\dagger}{\cal O}a^{(-)\dagger}}
e^{c^{(+)\dagger}\bar{c}^{(-)\dagger}
+c^{(-)\dagger}\bar{c}^{(+)\dagger}}\bar{c}_0
|p_{\mu}=0,x^i,\alpha\rangle\,\,.\label{eq:phiBa}
\end{eqnarray}
We claim that it satisfies the following relation
(``projector equation'' with the ghost insertion):
\begin{eqnarray}
&&|\Phi_B(\alpha_1) *\Phi_B(\alpha_2)\rangle 
=c_B |\Phi_B(\alpha_1+\alpha_2)\rangle\,,
~~~~~~~~~ \alpha_1 \alpha_2>0\,,
\label{eq:B2=B}
\\
&&
c_B\equiv V_{d-p-1}\,{\frak c}\,
\frac{\partial}{\partial \bar{c}_0}\,\,,
\\
&&  {\frak c}
\equiv e^{-2(\beta^2+\beta+1)\left({\log|\beta|\over \beta+1}
-{\log|\beta+1|\over \beta}\right)}
\left(\det(1-r^2)\right)^{-(d-2)/2}\,,~~~~~
\beta\equiv -\frac{\alpha_1}{\alpha_1+\alpha_2}\,.
\label{eq_frak_c}
\end{eqnarray}
$V_{d-p-1}$ is the volume of the Dirichlet directions.
The matrix $r$ is given by,
\begin{equation}
 r_{mn} = \frac{\beta(\beta+1)(mn)^{3/2}}{m+n}\bar{f}^{(3)}_m\bar{f}^{(3)}_n\,
, \quad
\bar{f}^{(3)}_m = \frac{\Gamma(-m\beta)
e^{m(\beta\log|\beta| -(\beta+1)\log|\beta+1|)}
}{m! \,\Gamma(-m\beta +1-m)}\,.
\end{equation}
${\frak c}$
depends on the ratio of $\alpha$ parameters.\footnote{
While we have not succeeded
in determining it analytically,
we can numerically evaluate it by truncating the
matrix $r$ to  $L\times L$.
We find that a good fit of this coefficient is
\begin{equation}
 \log({\frak c})\sim 3 \log(L) +7.07 + 0.866\,
(\beta^2+\beta+1)\left({\log|\beta|\over \beta+1}
-{\log|\beta+1|\over \beta}\right).
\end{equation}
At $L=100$, the error is about $\pm 0.02$.
This estimate shows that ${\frak c}/L^3$ is a finite and
well-behaved function of $\beta$. 
}

\item We consider the infinitesimal variations of $\Phi_B$ 
of the following form:\footnote{
Normal ordering which is necessary here is 
defined in appendix \ref{sec:Oscil Boundary}.
}
\begin{eqnarray}
 \delta_T |\Phi_B(\alpha)\rangle & = &
 \oint \frac{d\sigma}{2\pi} e^{ik_\mu \sqrt{\pi}X^\mu(\sigma)}
|\Phi_B(\alpha)\rangle\,,\label{eq_var_tachyon}\\
 \delta_V |\Phi_B(\alpha)\rangle & = &
 \oint \frac{d\sigma}{2\pi}
(\sqrt{\pi}\zeta_\nu \partial_\sigma X^\nu) 
e^{i\sqrt{\pi}k_\mu X^\mu(\sigma)}|\Phi_B(\alpha)\rangle\,\,.
\label{eq_var_vector}
\end{eqnarray}
The first (second) one corresponds to the tachyonic mode 
(vector particle) of the open string.
The infinitesimal variation of Eq.(\ref{eq:B2=B}),
\begin{equation}\label{eq:EOM}
 \delta |\Phi_B(\alpha_1)\rangle * |\Phi_B(\alpha_2)\rangle
+  |\Phi_B(\alpha_1)\rangle * \delta|\Phi_B(\alpha_2)\rangle
= c_B \delta|\Phi_B(\alpha_1+\alpha_2)\rangle
\end{equation}
gives the following constraints:
\begin{equation}\label{eq_on-shell}
 k_\mu G^{\mu\nu}k_\nu=2\,,\quad (\mbox{for }\delta=\delta_T)\,,
\,\quad
 k_\mu G^{\mu\nu}k_\nu=0\,,\quad (\mbox{for }\delta=\delta_V)\,,
\end{equation}
where
\begin{eqnarray}
 G^{\mu\nu}:=\left[{1+{\cal O}\over 2}{1+{\cal O}^T\over 2}\right]^{\mu\nu}
=\left[{(1+F)^{-1}}\eta{(1-F)^{-1}}\right]^{\mu\nu}
\end{eqnarray}
is the ``open string metric'' on the D$p$-brane.
These are precisely the mass-shell conditions for the tachyon
and the vector particles.  

The other part of the physical state conditions
for the vector particle,
the transversality condition $k_\nu G^{\nu\mu}\zeta_\mu=0$,
becomes rather subtle. At the level of the ``equation of motion'',
the coefficient of this factor takes the form $0\times \infty$
and we can not make a definite statement without more precise 
knowledge of the regularization scheme.

We note, however, that the variation (\ref{eq_var_vector})
is invariant under the gauge transformation;  namely, if we change
\begin{equation}
 \zeta_\mu \rightarrow \zeta_\mu + \epsilon k_\mu
\end{equation}
in Eq.(\ref{eq_var_vector}), $\delta_V |\Phi_B(\alpha)\rangle$
is not affected at all since the change can be written as the total
derivative with respect to $\sigma$ and it drops out after
the integration,
\begin{equation}
\oint \frac{d\sigma}{2\pi} (\sqrt{\pi}k_\nu \partial_\sigma X^\nu)
e^{ik_\mu \sqrt{\pi} X^\mu(\sigma)}|\Phi_B(\alpha)\rangle=
-i \oint \frac{d\sigma}{2\pi} \partial_\sigma
\left(e^{ik_\mu \sqrt{\pi}X^\mu(\sigma)}\right)|\Phi_B(\alpha)\rangle=0\,.
\label{eq:gaugeinv}
\end{equation}
In this sense, the gauge symmetry is automatically encoded in the 
vector particle.  
\end{enumerate}

One may give  an intuitive proof of
the projector equation Eq.(\ref{eq:B2=B}). We note that 
the boundary conditions (\ref{eq:BFi}--\ref{eq:BFpi}) 
for $|B(F)\rangle$ and the overlap conditions
(\ref{eq_over1}--\ref{eq_over3}) for $|V(1,2,3)\rangle$
are the local requirements on the boundary, namely they
are defined for each $\sigma$.  
Therefore, if we impose the same boundary conditions for
$|B_1\rangle$ and $|B_2\rangle$, they are translated into
the same boundary conditions for $|B_1 * B_2\rangle$
of the corresponding point.
Since the boundary state can be determined  from the boundary 
conditions up to the normalization, $|B_1 * B_2\rangle$
must be proportional to the same boundary state.
More explicit proof of this identity in terms of the Neumann coefficients
becomes, as we see below, rather lengthy 
while it is mostly straightforward. 
We have to use many
nontrivial identities of the Neumann coefficients.
In this sense the computation illuminates
a special r\^ole played by the boundary state.

\section{Proof of the idempotency of the boundary states
\label{sec:Gaussian}
}

In the following sections, we give the technical details
of the proof of Eqs.(\ref{eq:B2=B}, \ref{eq_on-shell}).
We first derive the star product of 
the boundary state which includes the additional linear term
in the exponential. It will be used to give the source
term to derive the variation of the boundary state.

We consider a tensor product of the boundary states,
\begin{eqnarray}
 \label{eq:Gaussian_Siegel}
&& |\Phi_1\rangle\otimes |\Phi_2\rangle
=e^{{1\over 2}a^{\dagger} M 
a^{\dagger}-\lambda a^{\dagger}} e^{-c^{\dagger}M_g \bar{c}^{\dagger}
}\bar{c}_0^{(1)}
|p_1,\alpha_1 \rangle\otimes \bar{c}_0^{(2)}|p_2,\alpha_2\rangle\,,
\end{eqnarray}
where we used abbreviated notation,
\begin{eqnarray}
&& a^\dagger=
\left(
\begin{array}{c}
 a^{(+)\dagger}\\a^{(-)\dagger}
\end{array}
\right)
\,,\quad
a^{(\pm)\dagger}=\left(
\begin{array}{c}
 a_n^{(1)(\pm)\mu\dagger}\\
 a_n^{(2)(\pm)\mu\dagger}
\end{array}
\right)
\,,\quad
\mbox{similar notation for $c,\bar{c}$}\,,\\
&&   M =\left(\begin{array}{c c}
     0 &      -\mathcal{O}_{MN} \delta_{mn}\delta_{rs}\\
     -\mathcal{O}^{T}_{MN} \delta_{mn}\delta_{rs} &  0
 \end{array}\right)\,,\quad
   M_g =\left(\begin{array}{c c}
     0 &     -\mathcal{O}_g \delta_{mn}\delta_{rs}\\
     -\mathcal{O}_g \delta_{mn}\delta_{rs} &   0
 \end{array}\right)\,,\label{eq_M}\quad\\
&&   \lambda =\left(\lambda^{(+)}\,,\,
\lambda^{(-)}\right)\,,\quad
\lambda^{(\pm)}=\left(
     \lambda_{n\,\mu}^{(1)(\pm)}\,,\,\, \lambda_{n\,\mu}^{(2)(\pm)}
\right)\,,\quad r,s=1,2\,;\quad m,n=1,2,\cdots,\infty\,.
\end{eqnarray}
We note that $\mathcal{O}_g=1$ for the conventional boundary state.
We include this extra degree of freedom since there exists
another choice $\mathcal{O}_g=-1$ which satisfies the projector
equation as we will see later.

The corresponding bra state is obtained by applying the reflector
and projectors,
\begin{eqnarray}
\langle \Phi_1|\wp\otimes
\langle \Phi_2|\wp
&=&\oint{d\theta_1\over 2\pi}{d\theta_2\over 2\pi}
\langle -p_1,-\alpha_1|{\bar{c}}_0^{(1)}\otimes 
\langle -p_2,-\alpha_2|{\bar{c}}_0^{(2)}\,
e^{{1\over 2}a M a+\lambda^{\theta} a}e^{cM_g \bar{c}} \,,
\end{eqnarray}
where\footnote{
The elements of $M,M_g$ do not change by $\wp$ 
because of the form (\ref{eq_M}).
}
\begin{eqnarray}
\lambda^{\theta}=\left(\lambda^{(+)\theta},\lambda^{(-)\theta}\right)\,,~~~~
\lambda^{(\pm)\theta}=\left(
e^{\mp in\theta_1}\lambda^{(1)(\pm)}_n\,,\,
e^{\mp in\theta_2}\lambda^{(2)(\pm)}_n
\right)\,.
\end{eqnarray}
We take the inner product between this state
with the 3-string vertex (\ref{eq:v123}).
For this purpose, it is convenient to rewrite the factor
in the exponential as
\begin{eqnarray}
 F(1,2,3) &=& \frac{1}{2} a^\dagger N a^\dagger  + a^\dagger\cdot\mu
  +a^{(3)\dagger} \tilde{N}^{33}a^{(3)\dagger}
-\frac{\tau_0}{4\alpha_1\alpha_2\alpha_3}{\bf P}^2
\nonumber\\
&& + c^\dagger N_g 
\bar{c}^\dagger + c^\dagger \cdot \rho + \sigma \cdot \bar{c}^\dagger\,
 + c_3^{(3)\dagger}
C^{1\over 2}\tilde{N}^{33}C^{-{1\over 2}}
 \bar{c}^{(3)\dagger} \,,
\end{eqnarray}
where we introduced some notation,
\begin{eqnarray}
&&N=\eta_{MN}\left(
\begin{array}{c c}
 n& 0 \\ 0 & n
\end{array}
\right)\,,\quad
n=\left(
\begin{array}{c c}
 \tilde{N}^{11}& \tilde{N}^{12} \\
 \tilde{N}^{21}& \tilde{N}^{22} 
\end{array}
\right)\,,\quad
N_g=\left(
\begin{array}{c c}
 n_g& 0 \\ 0 & n_g
\end{array}
\right)\,,\quad
n_g=\left(
\begin{array}{c c}
 \tilde{N}_g^{11}& \tilde{N}_g^{12} \\
 \tilde{N}_g^{21}& \tilde{N}_g^{22} 
\end{array}
\right)\,,~~~
\label{eq_def_n}\\
&&\tilde{N}^{rs}_{g} = \alpha_r C^{1\over 2}\tilde{N}^{rs}C^{-{1\over 2}}
\alpha_s^{-1}\,,\quad~~~
C_{mn}=m\delta_{mn}\,,
\\
&&
\mu =\left(
\begin{array}{c}
 \mu^{(+)}\\  \mu^{(-)}
\end{array}
\right)
\,,\quad
\mu^{(\pm)} = \left(
\begin{array}{c}
\tilde{N}^{13}a^{(3)(\pm)\dagger}
+{1\over 2}\tilde{N}^1{\bf P}
 \\
\tilde{N}^{23}a^{(3)(\pm)\dagger}
+{1\over 2}\tilde{N}^2{\bf P}
\end{array}
\right)\,,\quad  \\
&&
\rho=\left(
\begin{array}{c}
 \rho^{(+)}\\ \rho^{(-)}
\end{array}
\right)\,,\,~~
\rho^{(\pm)} = \left(
\begin{array}{c}
 \tilde N^{13}\bar c^{(3)(\pm)\dagger}_n\\
 \tilde N^{23}\bar c^{(3)(\pm)\dagger}_n
\end{array}
\right)\,,\\
&&
\sigma=\left(
 \sigma^{(+)}\,,\, \sigma^{(-)}
\right)\,,\,~~
\sigma^{(\pm)}=\left(
c^{(3)(\pm)\dagger}\tilde N_g^{31},\,c^{(3)(\pm)\dagger}\tilde N_g^{32}
\right)\,.
\end{eqnarray}
By taking the inner product with the aid of the useful formulae 
Eqs.(\ref{eq:gaussian_matter}),(\ref{eq:gaussian_ghost}),
in the appendix, we can arrive at the following
general formula after some calculation,\footnote{
In this expression and in the computation in the following,
we omit the suffix ${}^{(3)}$ in the oscillators.
The vector $a^\dagger$ 
should be interpreted as 
$(a_n^{\mu (3)(+)\dagger}\,,\,a_n^{\mu(3)(-)\dagger})^{T}$.
The same convention is also applied to  $c$ and $\bar{c}$.
}
\begin{eqnarray}
  |\Phi_1*\Phi_2\rangle&=& {\frak c}\wp
\oint{d\theta_1\over 2\pi}{d\theta_2\over 2\pi}
e^{H_{m}}\,{\cal{C}}\,e^{H_{g}}\bar c_0 |p_1+p_2,\alpha_1+\alpha_2\rangle\,\,,
\label{eq:Gaussian_star}\\
{\frak c}&=& [\mu(1,2,3)]^2\det{}^{-{1\over 2}}(1-MN)\det(1+N_gM_g)\,,
\\
H_m
&=&
{1\over 2}a^{\dagger}\tilde{N}^{33}a^{\dagger}
+{1\over 2}\tilde{N}^3(a^{(+)\dagger}+a^{(-)\dagger}){\bf P}
-{\tau_0\over 4\alpha_1\alpha_2\alpha_3}{\bf P}^2
\nonumber  \\
&&+{1\over 2}\mu M(1-NM)^{-1}\mu+\lambda^{\theta}(1-NM)^{-1}\mu
+{1\over 2}\lambda^{\theta} N(1-MN)^{-1}\lambda^{\theta}\,,\label{eq_Hm}\\
H_g&=&c^{\dagger}C^{1\over 2}\tilde{N}^{33}C^{-{1\over 2}}
\bar{c}^{\dagger}-\sigma(1+M_gN_g)^{-1}M_g\rho\,
 \,,
\label{eq_Hg}\\
{\mathcal C}&=&{\partial \over \partial \bar{c}_0}
+{1\over 2}\sum_{n=1}^{\infty}\alpha_3(c^{(+)\dagger}_n+c^{(-)\dagger}_n)
\nonumber\\
&&~~~~~~~~~~~~~~\times\left(w^{33}_n
+\sum_{r,s=1,2}\left(C^{1\over 2}\tilde{N}^{3r}C^{-{1\over 2}}\alpha_r^{-1}
((1-{\cal O}_gN_g)^{-1}{\cal O}_g)^{rs}\alpha_sw^{3s}\right)_n
\right)\,.
\end{eqnarray}
In the derivation of this formula, we do not use the information
of the particular form of $M$.  In this sense, this gives
the general formula for the star product of the
generic squeezed states 
of the form (\ref{eq:Gaussian_Siegel}).

This expression looks hopelessly complicated.
In particular, the appearance of the inverse of Neumann coefficients,
$(1-NM)^{-1}$ or $(1+N_g M_g)^{-1}$
in $H_m$ and $H_g$, looks unmanageable and even singular
for generic $M$.

A major simplification occurs, however,
 when we replace the matrices $M,M_g$ with
those of the form (\ref{eq_M}).
In this case, one may use
\begin{equation}
 (1-NM)^{-1} = \left(
\begin{array}{c c}
 1& n\, \mathcal{O}\\  n \, \mathcal{O}^T &1
\end{array}
\right)^{-1}=\left(
\begin{array}{c c}
 (1-n^2)^{-1}& -\mathcal{O}\, n(1-n^2)^{-1} \\
 -\mathcal{O}^{T}\, n(1-n^2)^{-1} & (1-n^2)^{-1}
\end{array}
\right)\,,
\end{equation}
and similar one for the ghost sector.
We note that $\mathcal{O}$ and $n$ commute with each other since
the matrix $\mathcal{O}$ acts on the Lorentz
indices while the Neumann coefficients acts only the level index.
The problem is reduced to deriving the inverse $(1-n^2)^{-1}$.
At first look, this is singular since the relation
(\ref{eq:yoneya}) among Neumann coefficients implies
\begin{equation}
(1-n^2)^{rs}_{nm}=
\sum_{\ell}\tilde{N}^{r3}_{n\ell}\tilde{N}^{3s}_{\ell m}~~~~~
(r,s=1,2). \label{eq:1-n2=NN}
\end{equation}
On the left hand side, the size of the matrix 
with respect to the indices $\{(r,n),(s,m)\}$
is ``$2\times \infty$''
whereas the summation on the right hand side is taken over ``$\infty$''
set. 
If we naively regularize the Neumann matrices
 $\tilde{N}^{r3},\tilde{N}^{3s}\,(r,s=1,2)$ by truncating their
 size to $L$ respectively, 
the rank of $(1-n^2)$ becomes $L$ while its size is $2L$.

It is a surprise that, contrary to this
naive expectation, it has a well-defined inverse.
This is a specialty of the infinite dimensional
matrices. 
For the explicit computation, we need detailed forms of
the Neumann coefficients
\begin{eqnarray}
 \tilde{N}^{rs}_{mn} &=& \delta_{rs}\delta_{mn} -
2 (A^{(r)T} \Gamma^{-1}A^{(s)})_{mn}\,,\quad
\tilde{N}^r_m = -(A^{(r)}\Gamma^{-1}B)_m\,,\\
A^{~(1)}_{mn} &=& -\frac{2}{\pi}\sqrt{mn}(-1)^{m+n}
\frac{\beta \sin(m\pi\beta)}{n^2-m^2\beta^2}\,,\quad
A^{~(2)}_{mn} = -\frac{2}{\pi}\sqrt{mn}(-1)^{m}
\frac{(\beta+1) \sin(m\pi\beta)}{n^2-m^2(\beta+1)^2}\,,
\\
A^{(3)}_{mn}&=&\delta_{mn}\,,\quad
B_m=-\frac{2\alpha_3}{\pi \alpha_1\alpha_2} m^{-3/2}(-1)^m 
\sin(m\pi\beta)\,,\quad
\Gamma_{mn}=\delta_{mn}+\sum_{r=1,2} (A^{(r)}A^{(r)T})_{mn}\,\,.
\end{eqnarray}
$A^{(r)}$ and $B$ describe the overlap of Fourier basis
of three strings at the vertex.  A crucial property of $A^{(r)}$ ($r=1,2$)
is that they have an inverse, which was proved
in \cite{GS},
\begin{equation}\label{eq_def_D}
\sum_{r=1,2}A^{(r)}D^{(r)}=1\,,\quad D^{(r)}A^{(s)}=\delta_{rs}\,,\quad
 D_{mn}^{(r)}\equiv -\frac{\alpha_3}{\alpha_r} 
CA^{(r)T}C^{-1}\quad(r,s=1,2)\,.
\end{equation}
By using this inverse, one obtains the inverse of $1-n^2$,
\begin{equation}
 (1-n^2)^{rs}= 4A^{(r)T}\Gamma^{-2}A^{(s)}\quad \Longrightarrow\quad
((1-n^2)^{-1})^{rs}= \frac{1}{4}D^{(r)}\Gamma^2 D^{(s)T}\,.
\end{equation}
With this remark, we derive the following relations which
are essential to show the idempotency of the boundary state,
\begin{eqnarray}
 && \tilde N^{33}+ \sum_{r,s,t=1,2}\tilde N^{3r}n^{rt}((1-n^2)^{-1})^{ts}
\tilde N^{s3}=0\,,\quad
\sum_{r,s=1,2} \tilde N^{3r}((1-n^2)^{-1})^{rs} \tilde N^{s3}=1\,.
\label{eq_Neum1}
\end{eqnarray}
We list many other formulae for Neumann coefficients in appendix
\ref{sec:formulae}.

In the next section, we will need cut off the range of  the lower indices 
of the Neumann coefficients to obtain a finite result.
We need impose the condition that $\sum_\ell 
\tilde N^{r3}_{p\ell} \tilde N^{3s}_{\ell q}$ 
(\ref{eq:1-n2=NN})
has the inverse as mentioned above even
in the finite size truncation.
This observation will have an important consequence later.

Now we come back to the computation of $*$ product.
By using the relations (\ref{eq_Neum1}), we can simplify
the gaussian part of Eq.(\ref{eq:Gaussian_star}).
We neglect the $\lambda$ dependence for the moment
since they are not relevant in the proof of the idempotency
of the boundary states. The exponents \eqref{eq_Hm} and \eqref{eq_Hg} are
\begin{eqnarray}
H_m &=&
{1\over 2}a^{\dagger}M a^{\dagger}
-{1\over 2}\left(a^{(+)\dagger}{1+{\cal O}\over 2}+a^{(-)\dagger}
{1+{\cal O}^T\over 2}\right)B{\bf P}
-{1\over 8}{\bf P}{1+{\cal O}\over 2}{\bf P}B^T B\nonumber\\
&&~~~+{1\over 2}\mu^T{N\over (1-NM)(1-N^2)}(M^2-1)\mu\,,
\end{eqnarray}
for the matter sector and
\begin{eqnarray}
H_g&=&
-c^{\dagger}M_g\bar{c}^{\dagger}
+({\cal O}_g^2-1)\,c^{\dagger}C^{1\over 2}\tilde{N}^{3r}_g\left(
{N_g\over (1+M_g N_g)(1-N^2_g)}
\right)^{rs}\tilde{N}^{s3}_gC^{-{1\over 2}}\bar{c}^{\dagger}
\end{eqnarray}
for the ghost sector. These expressions are further
simplified 
by the following conditions
\begin{eqnarray}
 &&M^2=1\,,~M_g^2=1\,,~~
{1+{\cal O}^T\over 2}{\bf P}={1+{\cal O}\over 2}{\bf P}=0
\label{eq:unipotency}
\\
&\leftrightarrow&~~~
{\cal O}^T{\cal O}=1\,,~~~{\cal O}_g=\pm 1\,,~~~
{1+{\cal O}\over 2}(\alpha_1p_2-\alpha_2p_1)=0
\label{eq:M2=1}
\end{eqnarray}
which are satisfied automatically for the conventional boundary state
(\ref{eq:phiBa}).
After we use these relations,
we arrive at the final result,
\begin{equation}
 H_m+H_g = \frac{1}{2}a^\dagger M a^\dagger 
-c^\dagger M_g \bar{c}^\dagger\,.
\end{equation}
We note that commutativity: $[M,N]=[M_g,N_g]=0$ 
which follows from Eq.(\ref{eq_M}) and unipotency of $M,M_g$ 
(\ref{eq:unipotency}) are sufficient conditions 
to derive this form.

The ghost prefactor ${\cal C}$ in Eq.(\ref{eq:Gaussian_star}) becomes 
\begin{equation}
 {\cal C}={\partial\over \partial\bar{c}_0}+{1\over 2}\alpha_3
\left(w^{33}+w^{3s}{\cal O}_gC^{-{1\over 2}}((1-{\cal O}_g n)^{-1})^{sr}
\tilde{N}^{r3}C^{1\over 2}\right)(c^{(+)\dagger}+c^{(-)\dagger})\,.
\label{eq:calC}
\end{equation}
It is simplified further for ${\cal O}_g=+1$,
\begin{eqnarray}
 {\cal C}_+&=&{\partial\over \partial\bar{c}_0}+{1\over 2}\alpha_3
\left(w^{33}-\sum_{r=1}^2
w^{3r}C^{-{1\over 2}}D^{(r)}C^{1\over 2}
\right)(c^{(+)\dagger}+c^{(-)\dagger})
= \frac{\partial}{\partial\bar c_0}\,,
\label{eq_C+}
\end{eqnarray}
and for ${\cal O}_g=-1$,
\begin{eqnarray}
 {\cal C}_-&=&{\partial\over \partial\bar{c}_0}+{1\over 2}\alpha_3
\left(w^{33}+\sum_{r=1}^2
w^{3r}C^{-{1\over 2}}A^{(r)T}C^{{1\over 2}}
\right)(c^{(+)\dagger}+c^{(-)\dagger})\\
&=& \frac{\partial}{\partial \bar c_0} +
\sum_{n=1}^\infty \cos(n\pi(\beta+1))(c^{(+)\dagger}_n+c^{(-)\dagger}_n)\,.
\label{eq_C-}
\end{eqnarray}
At the final stage in Eqs.(\ref{eq_C+},\ref{eq_C-}),
we have used relations \cite{HIKKO1},
\begin{eqnarray}
 w^{rs}_m & = & m\left(
\chi^{rs}\bar{N}^s_m +\frac{1}{\alpha_r}\sum_{n=1}^{m-1}
\bar{N}^{ss}_{m-n,n}
\right)
= \frac{1}{\alpha_r}\left(
\delta_{r,s} \cos m\sigma_r
-m\sum_{n=1}^\infty \bar{N}_{mn}^{sr}\cos n\sigma_{r}
\right)\,,~~(r,s=1,2,3)\nonumber\\
&&~~~~~ {\textrm{where}}~~~~\sigma_1=\pi\,,\quad
\sigma_2=0\,,\quad
\sigma_3=\pi(\beta+1)\,.
\end{eqnarray}
After computing,  the normalization factor
${\frak c}$ is 
\begin{eqnarray}
{\frak c}&= &
[\mu(1,2,3)]^2(\det(1-n^2))^{-{d-2\over 2}}\nonumber\\
&=&e^{-2(\beta^2+\beta+1)\left({\log|\beta|\over \beta+1}
-{\log|\beta+1|\over \beta}\right)}
\left(\det(4A^{(r)T}\Gamma^{-2}A^{(s)})_{r,s=1,2}\right)^{-(d-2)/2}\,.
\end{eqnarray}
The second factor is simplified by
$
 \det(4(\sum_{r=1}^2A^{(r)}A^{(r)T})\Gamma^{-2} )
= \det(4(\Gamma-1)\Gamma^{-2})
$.
After we use the expression $\tilde N^{33}=1-2\Gamma^{-1}$ and 
the relations (\ref{eq_Neum2}--\ref{eq_Neum4}) in appendix
\ref{sec:formulae}, we obtain Eq.(\ref{eq_frak_c}).

Thus far, 
for the string field
\begin{equation}
\label{eq:Phi0(pa)}
 |\Phi_0(p,\alpha)\rangle=e^{-a^{(+)\dagger}{\cal O}a^{(-)\dagger}
+{\cal O}_g (c^{(+)\dagger}\bar{c}^{(-)\dagger}
+c^{(-)\dagger}\bar{c}^{(+)\dagger})
}\bar{c}_0|p,\alpha\rangle\,,
~~~~{\cal O}^T{\cal O}=1\,,~{\cal O}_g=\pm 1\,,
\end{equation}
we have derived
\begin{eqnarray}
 |\Phi_0(p_1,\alpha_1) *\Phi_0(p_2,\alpha_2)\rangle 
={\frak c}\wp\,{\cal C}_{\pm}|\Phi_0(p_1+p_2,\alpha_1+\alpha_2)\rangle
\label{eq:Phi2=Phi}
\end{eqnarray}
with nonvanishing momentum only along Dirichlet directions\footnote{
The condition for momentum comes from ${1+{\cal O}\over
2}(\alpha_1p_2-\alpha_2p_1)=0$ in Eq.(\ref{eq:M2=1}).}
where ${\frak c},{\cal C}_{\pm}$ are given by
Eqs. (\ref{eq_frak_c}), (\ref{eq_C+}), (\ref{eq_C-}), respectively.
Note that in Eq. \eqref{eq:Phi2=Phi}, the ghost prefactor ${\cal C}_-$
can be regarded as ${\cal C}_+$ since the second term in Eq.\eqref{eq_C-}
vanishes after projection by $\wp$.

In  the case of ${\cal O}_g=+1$,
the final result Eq.(\ref{eq:B2=B}) with the volume factor
is obtained
by Fourier transformation along the Dirichlet
direction ($x^i$)
\begin{eqnarray}
 &&|x^i\rangle=\int {d^{d-p-1}p\over (2\pi)^{d-p-1}} e^{-i p_i x^i}
|p_i\rangle\,,\\
&& \int {d^{d-p-1}p_1\over (2\pi)^{d-p-1}}{d^{d-p-1}p_2\over (2\pi)^{d-p-1}}
{d^{d-p-1}p_3\over (2\pi)^{d-p-1}}
{}_1\langle x^i_1| {}_2\langle x^i_2|
(2\pi)^{d-p-1}\delta^{d-p-1}(p_1+p_2+p_3)
|p_1\rangle_1|p_2\rangle_2|p_3\rangle_3\nonumber\\
&&~~~=\int{d^{d-p-1}p_2\over (2\pi)^{d-p-1}}e^{i(x_2-x_1)^ip_{2i}}
\int{d^{d-p-1}p\over (2\pi)^{d-p-1}}e^{ip_ix^i_1}|-p\rangle_3
=\delta^{d-p-1} (x_1 - x_2)\,|x^i_1\rangle_3
\,.
\end{eqnarray}
In the case of $x_1^i = x_2^i$ as Eq.(\ref{eq:B2=B}),
we set the divergent coefficient 
$\delta^{d-p-1}(0)$ in the last line as $V_{d-p-1}$.

Finally we make a few comments on the analogy with
VSFT. We note that there exists an extra solution of idempotency relation, 
the $|\Phi_0\rangle$ 
(\ref{eq:Phi0(pa)}) with ${\cal O}_g=-1$.
The ghost prefactor ${\cal C}_-$ can be rewritten as
\begin{equation}
 {\cal C}_-|\Phi_0\rangle={\sqrt{\pi}\over 2}
\left(i\pi_{\bar{c}}(\pi(\beta+1))+i\pi_{\bar{c}}(-\pi(\beta+1))\right)
|\Phi_0\rangle\,,
\end{equation}
where $\sigma_3=\pm \pi(\beta+1)$ are the interaction points of
the string $r=3$.
It has a similar form to BRST operator
in VSFT \cite{GRSZ}: $c(\pi/2)$ in Witten-type 
open string field theory; 
namely, if we regard the ``projector equation'' (\ref{eq:Phi2=Phi}) as
 an analogue of the equation of motion of VSFT,
the string field $|\Phi_0\rangle$ with  ${\cal O}_g=-1$ corresponds to 
Hata-Kawano's ``sliver-like'' solution of VSFT.\cite{Okuyama:2002yr}
On the other hand, in the case of ${\cal O}_g=+1$, $|\Phi_0\rangle$ 
 corresponds to
``identity-like'' solution of VSFT \cite{IK} with the same analogy
of ${\cal C}_+={\partial\over\partial\bar{c}_0}\sim c_0$.
Although there are two choices for ghost sector: 
${\cal O}_g=\pm 1$ for Eq.(\ref{eq:Phi2=Phi}),
only $|\Phi_0\rangle$ with ${\cal O}_g=+1$ relates
 to the boundary states (\ref{eq:B(F)})
which have conventional BRST invariance.
In the following section, we discuss only $|\Phi_B\rangle$ (\ref{eq:phiBa})
with ${\cal O}_g=+1$.

\section{Fluctuation around projectors
\label{sec:fluctuation}
}
In this section we consider two types of fluctuations 
Eqs.(\ref{eq_var_tachyon},\ref{eq_var_vector})
around $|\Phi_B \rangle$ and demonstrate explicitly 
that the idempotency condition (\ref{eq:EOM}) produces
the on-shell conditions (\ref{eq_on-shell}) for these particles.
We note that the variations of the type 
\begin{equation}
 |\delta \Phi_B\rangle = \oint \frac{d\sigma}{2\pi}
 V(\sigma) |\Phi_B\rangle\,,
\end{equation}
correspond to the open string modes on the D-brane (see, for example
 \cite{Callan}).
We conjecture that the ``equation of motion'' Eq.\eqref{eq:Phi2=Phi}
will produce
the on-shell condition for all of them, namely, they should
be the marginal deformation on the boundary.
We pick the simplest two examples to illustrate this idea
explicitly.

Before we start the computation for these cases, 
we give a few technical remarks.
\begin{enumerate}
 \item By using Eq.(\ref{eq:Gaussian_star}) with nonzero $\lambda$
as the generating functional, 
we will compute the left hand side of Eq.(\ref{eq:EOM}).
Explicitly, for a fluctuation
\begin{eqnarray}
 |\delta \Phi_B\rangle=
a^{(\pm)\dagger}_{n_1}\cdots a^{(\pm)\dagger}_{n_l}e^{-\lambda_0a^{\dagger}}
|\Phi_B\rangle \,,
\end{eqnarray}
we can compute $*$ product as
\begin{eqnarray}
 |\delta \Phi_B*\Phi_B\rangle=\left[
\left(-{\partial\over \partial\lambda_{n_1}^{\pm}}\right)
\cdots \left(-{\partial\over \partial\lambda_{n_l}^{\pm}}\right)
|(e^{-\lambda a^{\dagger}}\Phi_B)*\Phi_B\rangle
\right]_{\lambda=\lambda_0}\,,\label{eq:del_lambda}
\end{eqnarray}
and $|\Phi_B*\delta \Phi_B\rangle$ similarly.

\item From the definition of the 
tensor product of the  boundary states,
Eq.(\ref{eq:Gaussian_Siegel}), we can define
projection matrices ${\cal P}_{\pm}$ as\footnote{
$M$ in the following should be interpreted as
Eq.(\ref{eq_M}) with $\delta_{rs}$ dropped.}
\begin{eqnarray}
 {\cal P}_{\pm}={1\mp M\over 2}\label{eq:Ppm}
\end{eqnarray}
which satisfy
\begin{eqnarray}
 &&{\cal P}_{\pm}^2={\cal P}_{\pm}\,,~~~
{\cal P}_{\pm}^T={\cal P}_{\pm}\,,~~~
{\cal P}_{\pm}{\cal P}_{\mp}=0\,,~~~
{\cal P}_++{\cal P}_-=1\,,
\end{eqnarray}
because of Eqs.(\ref{eq:M2=1}).
These projection operators are useful for classifying the
external source term into odd and/or even parts 
under the reflection at the boundary and 
for simplifying the computations.
\item $\lambda$-dependent terms of the exponent in
Eq.(\ref{eq:Gaussian_star}) can be simplified by using
the identities of Neumann coefficients
(appendix \ref{sec:list-neumann}),\footnote{
We denote 
$\lambda^{(r)\theta_r}=
(\lambda^{(r)(+)\theta_r},\,\lambda^{(r)(-)\theta_r})
=(e^{-in\theta_r}\lambda^{(r)(+)}_n,\,e^{in\theta_r}\lambda^{(r)(-)}_n)
\,,r=1,2$.
}
\begin{eqnarray}
&&{1\over 2}\lambda^{\theta} N(1-MN)^{-1}\lambda^{\theta}=
{1\over 4}\sum_{r,s=1}^2\lambda^{(r)\theta_r}{\cal P}_+\left(
\delta_{r,s}-A^{(r)T}A^{(s)}
\right){\cal P}_+\lambda^{(s)\theta_s} \nonumber\\
&&~~~~~~~~~~~~~~~~~~~~~~~~~~
-{1\over 4}\sum_{r,s=1}^2\lambda^{(r)\theta_r}{\cal P}_-\left(
\delta_{r,s}-D^{(r)}D^{(s)T}
\right){\cal P}_-\lambda^{(s)\theta_s},
\label{eq:lambda2}\\
&&\lambda^{\theta}(1-NM)^{-1}\mu=-\sum_{r=1}^2
\lambda^{(r)\theta_r}{\cal P}_+A^{(r)T}{\cal P}_+a^{\dagger}
-\sum_{r=1}^2\lambda^{(r)\theta_r}{\cal P}_-D^{(r)}{\cal P}_-a^{\dagger}
\nonumber\\
&&~~~~~~~~~~~~
-{1\over 4}\sum_{r=1}^2\lambda^{(r)\theta_r}{\cal P}_+A^{(r)T}B
\left(
\begin{array}[tb]{c}
{1+{\cal O}\over 2} \\
{1+{\cal O}^T\over 2}
\end{array}
\right){\bf P}
-{1\over 4}\sum_{r=1}^2\lambda^{(r)\theta_r}{\cal P}_-D^{(r)}B\left(
\begin{array}[tb]{c}
{1-{\cal O}\over 2} \\
{1-{\cal O}^T\over 2}
\end{array}
\right){\bf P}\,.\label{eq:lambdaa}
\end{eqnarray}
\end{enumerate}

\subsection{Tachyon type fluctuation}

We consider tachyon type fluctuation of the form
Eq.(\ref{eq_var_tachyon}).
After we use the identification of the oscillators 
on the boundary state (\ref{eq_b_a1}--\ref{eq_b_c}),(\ref{eq:V_TB}),
the variation takes the following form,
\begin{eqnarray}
 |\delta_T \Phi_B(\alpha)\rangle
&=&\oint {d\sigma\over 2\pi}V_T(\sigma)|\Phi_B(\alpha)\rangle\nonumber\\
&=&e^{ik_ix^i}\oint {d\sigma\over 2\pi}
e^{-a^{(+)\dagger}{\cal O}a^{(-)\dagger}-\lambda^{\sigma}
 a^{\dagger}}
e^{c^{(+)\dagger}\bar{c}^{(-)\dagger}
+c^{(-)\dagger}\bar{c}^{(+)\dagger}}\bar{c}_0|k_{\mu},x^i,\alpha
\rangle\,,\\
\lambda^{\sigma}_{\nu}&=&-k_{\mu}\left(
\left({1+{\cal O}^T\over 2}\right)^{\mu}_{~\nu}
{e^{-in\sigma}\over \sqrt{n}}\,,~~
\left({1+{\cal O}\over 2}\right)^{\mu}_{~\nu}
{e^{in\sigma}\over \sqrt{n}}\right)\nonumber\\
&=&-k(\cos n\sigma)C^{-{1\over2}}(1,1){\cal P}_+
+i k(\sin n\sigma)C^{-{1\over2}}(1,-1){\cal P}_-\,,\label{eq:lambda^sigma}
\\
\lambda^{\sigma}_i&=&0\,.
\end{eqnarray}
We need some explanations on our notation.
The bra vector $(\cos n\sigma)$ (or $(\sin n\sigma)$)
has only the level index $n$ and whose $n$-th component
is $\cos n\sigma$ (or $\sin n\sigma$).
In this notation, we may also write, for example,
$(\cos(n\sigma)/\sqrt{n})\equiv (\cos(n\sigma))C^{-1/2}$ and so on.
The other bra  vector $(1\,,\, 1)$ has the index $\pm$ 
which distinguishes the left and right movers.
Finally ${\cal P}_{\pm}$ have the indices of Lorentz and
left/right $\pm$ as was defined in Eqs.(\ref{eq:Ppm}),(\ref{eq_M}).

The integration with respect to $\sigma$ 
appears automatically
because of the  projection $\wp$ in the 
definition of the $*$ product (\ref{eq:Gaussian_star}).

We investigate ``on-shell'' condition which is imposed by
Eq.(\ref{eq:EOM}). We evaluate  the $*$ products
\begin{eqnarray}
 |(V_T(\sigma)\Phi_B(\alpha_1))*\Phi_B(\alpha_2)\rangle
&=&c_B\,\wp e^{ik_ix^i}\oint {d\theta_1\over 2\pi}
e^{E_1}e^{c^{(+)\dagger}\bar{c}^{(-)\dagger}
+c^{(-)\dagger}\bar{c}^{(+)\dagger}}\bar{c}_0
|k_{\mu},x^i,\alpha_1+\alpha_2\rangle\,,\label{eq:VBB}\\
 |\Phi_B(\alpha_1)*(V_T(\sigma)\Phi_B(\alpha_2))\rangle
&=&c_B\, \wp e^{ik_ix^i}\oint {d\theta_2\over 2\pi}
e^{E_2}e^{c^{(+)\dagger}\bar{c}^{(-)\dagger}
+c^{(-)\dagger}\bar{c}^{(+)\dagger}}\bar{c}_0
|k_{\mu},x^i,\alpha_1+\alpha_2\rangle\,.\label{eq:BVB}
\end{eqnarray}
The calculation of $E_1, E_2$ is reduced to 
that of $H_m$ in the previous section,
Eq.(\ref{eq_Hm}).
We have already evaluated first four terms
while we need to keep the nontrivial $k$ dependence.
The last two terms are simplified in
Eqs.(\ref{eq:lambda2}),(\ref{eq:lambdaa}).
In the computation of $E_1$ we put $\lambda^{(2)}=0$.
\begin{eqnarray}
E_1
&=&E_1^{[2]}+E_1^{[1]}+E_1^{[0]}\,,\\
E_1^{[2]} & = & -a^{(+)\dagger}{\cal O}a^{(-)\dagger}\,,
\\
E_1^{[1]} &=& 
-{1\over 2}\left(a^{(+)\dagger}{1+{\cal O}\over 2}+a^{(-)\dagger}
{1+{\cal O}^T\over 2}\right)\alpha_2Bk -\lambda^{\sigma+\theta_1}{\cal P}_+A^{(1)T}{\cal P}_+a^{\dagger}
-\lambda^{\sigma+\theta_1}{\cal P}_-D^{(1)}{\cal P}_-a^{\dagger}\,,
\\
E_1^{[0]}&=&-{\alpha_2^2\over 8}B^TB\,k{1+{\cal O}\over 2}k
-{1\over 4}\lambda^{\sigma+\theta_1}{\cal P}_+A^{(1)T}\alpha_2 B
\left(
\begin{array}[tb]{c}
{1+{\cal O}\over 2} \\
{1+{\cal O}^T\over 2}
\end{array}
\right)k
-{1\over 4}\lambda^{\sigma+\theta_1}{\cal P}_-
D^{(1)}\alpha_2 B\left(
\begin{array}[tb]{c}
{1-{\cal O}\over 2} \\
{1-{\cal O}^T\over 2}
\end{array}
\right)k
\nonumber\\
&&+{1\over 4}\lambda^{\sigma+\theta_1}{\cal P}_+\left(
1-A^{(1)T}A^{(1)}\right){\cal P}_+\lambda^{\sigma+\theta_1}
-{1\over 4}\lambda^{\sigma+\theta_1}{\cal P}_-\left(
1-D^{(1)}D^{(1)T}
\right){\cal P}_-\lambda^{\sigma+\theta_1}\,.
\end{eqnarray}
The quadratic part in the oscillator $E^{[2]}_1$
is the same as the boundary state.
To calculate $E^{[1]}_1$ and $E^{[0]}_1$, 
we need to evaluate the inner products
between the vectors
$(\cos n\sigma)C^{-1/2}$ or $(\sin n\sigma)C^{-1/2}$
with matrices $A^{T(r)}$ and $D^{(r)}$.  They are reduced to the calculation
of Fourier transformation which we explain in detail in
Appendix \ref{sec:formulae}, Eqs.(\ref{eq:sum_cos}--\ref{eq:cosD2}).
They simplify the linear part dramatically to
\begin{equation}
 E^{[1]}_1 = -\lambda^{-\beta(\sigma+\theta_1)+\pi} a^\dagger\,.
\end{equation}
This is identical to the linear part coming from the
tachyon vertex.  The constant part is similarly computed, as follows:
\begin{eqnarray}
E^{[0]}_1&=&
{1\over 2}k{1+{\cal O}\over 2}k
\biggl[
-{\alpha_2^2\over 4}B^TB
+{4\over \pi^2}\sum_{m=1}^{\infty}\left({\sin^2 m\pi\beta\over 2m^3\beta^2}
-{\pi\over 2m^2\beta}\sin m\pi\beta \cos m\beta(\sigma+\theta_1)
\right)\nonumber\\
&&+\sum_{m,n=1}^{\infty}
\cos m(\sigma+\theta_1)\cos n(\sigma+\theta_1)
\left({\delta_{m,n}\over m}-{4\beta^2\over \pi^2}(-1)^{m+n}
\sum_{p=1}^{\infty}{p\sin^2p\pi\beta\over (m^2-p^2\beta^2)(n^2-p^2\beta^2)}
\right)\label{eq:t-mass}         \\
&&+\sum_{m,n=1}^{\infty}
\sin m(\sigma+\theta_1)\sin n(\sigma+\theta_1)
\left({\delta_{m,n}\over m}-{4mn(-1)^{m+n}\over \pi^2}
\sum_{p=1}^{\infty}{\sin^2p\pi\beta\over p(m^2-p^2\beta^2)(n^2-p^2\beta^2)}
\right)\biggr]\,.\nonumber
\end{eqnarray}
The overall factor of $E^{[0]}_1$ becomes
\begin{equation}
 \frac{1}{2}k\frac{1+\mathcal{O}}{2} k= \frac{1}{2}
k_\mu G^{\mu\nu}
k_{\nu}\,,\quad
 G^{\mu\nu}:=\left[{1+{\cal O}\over 2}{1+{\cal O}^T\over 2}\right]^{\mu\nu}
=\left[{(1+F)^{-1}}\eta{(1-F)^{-1}}\right]^{\mu\nu}\,.
\end{equation}
$G^{\mu\nu}$ is ``open string metric'' on the D$p$-brane.

We evaluate the numerical factor  $\left[\cdots\right]$ 
in Eq.(\ref{eq:t-mass}).
The quantities in the first line are convergent.
On the other hand,
the evaluation of the terms in the second and third lines
are very subtle.
Two terms with $\delta_{mn}$ can be summed to give
$\sum_{m=1}^\infty 1/m$ which diverges logarithmically.
The summation of the other two terms,
if we first perform $\sum_{m,n=1}^{\infty}$
using Eqs.(\ref{eq:sum_cos}),(\ref{eq:sum_sin}), 
gives again $-\sum_{p=1}^\infty 1/p$, which is
divergent  but with negative sign:
\begin{eqnarray}
 &&\sum_{m,n=1}^{\infty}
\cos m(\sigma+\theta_1)\cos n(\sigma+\theta_1)
\left(-{4\beta^2\over \pi^2}(-1)^{m+n}
\sum_{p=1}^{\infty}{p\sin^2p\pi\beta\over (m^2-p^2\beta^2)(n^2-p^2\beta^2)}
\right)\nonumber\\
&&+\sum_{m,n=1}^{\infty}
\sin m(\sigma+\theta_1)\sin n(\sigma+\theta_1)
\left(-{4mn(-1)^{m+n}\over \pi^2}
\sum_{p=1}^{\infty}{\sin^2p\pi\beta\over p(m^2-p^2\beta^2)(n^2-p^2\beta^2)}
\right)\nonumber\\
&=&-{4\over \pi^2}\sum_{p=1}^{\infty}\sin^2p\pi\beta
\left(p\left({1\over 2p^2\beta^2}-{\pi\cos p\beta(\sigma+\theta_1)
\over 2p\beta\sin p\pi\beta}\right)^2
+{1\over p}\left({\pi\sin p\beta(\sigma+\theta_1)
\over 2\sin p\pi\beta}\right)^2
\right)\nonumber\\
&=&\sum_{p=1}^{\infty}\left(-{\sin^2 p\pi\beta\over \pi^2p^3\beta^2}
+{2\sin p\pi\beta\cos p\beta(\sigma+\theta_1)\over\pi p^2\beta}
-{1\over p}\right)\,.
\end{eqnarray}
The summation of the first two terms are finite and 
exactly cancel with the first line of Eq.(\ref{eq:t-mass}).
As for the third term, 
we encounter subtle cancellation of the form
$
[\cdots]=\sum_{m=1}^{\infty}{1\over m}-\sum_{p=1}^{\infty}{1\over p}=
\infty-\infty\,.$
We need some regularization to obtain a finite result.\footnote{
There was a similar subtlety of the tachyon mass around the sliver solution
in the oscillator approach of VSFT which was proposed in \cite{HK}.
As was shown in \cite{HM},\cite{HMT}, the correct mass was reproduced
using a regularization of the Neumann matrices
although it becomes divergent if one uses relations among them naively.
}
For this purpose, we cut off
the infinite dimensional matrix for $A^{(r)}$.
As we commented in the previous section, in order that
$A^{(r)}$ has the inverse $D^{(r)}$
in the sense of Eq.(\ref{eq_def_D}), we should regard
$A^{~(r)}_{pm}~(r=1,2)$ (resp. $D^{(r)}_{mp}$) as sub-blocks of an infinite
dimensional square matrix 
$A\equiv \left(A^{(1)}, A^{(2)}\right)$ (resp. $D=\left( 
\begin{array}{c}D^{(1)}\\ D^{(2)}\end{array} 
\right)$).  With these combinations, the relation Eq.(\ref{eq_def_D}) becomes
simply $AD=DA=1$. In the cut-off regularization, we demand
$A, D$ to become $L\times L$ matrices with a large integer $L$. 
It implies that the sub-blocks $A^{(r)}$  become rectangular
matrix with size $L\times L_r$ with $L_1+L_2=L$.
In the following, we demand
\begin{equation}
L_1: L_2: L\rightarrow \alpha_1 : \alpha_2 : (\alpha_1+\alpha_2) =
(-\beta) : (1+\beta) : 1\,,
\quad \mbox{as}\ L\rightarrow \infty.
\end{equation}
%
An explanation of this division is
to  come back to the definitions of $A^{(r)}$ summarized in the appendix (C.1).
The first lower index $p$ (resp. the second lower  $m$) in $A^{~(r)}_{pm}$ 
labels the Fourier bases $\cos(p\sigma/\alpha_3)$ 
(resp. $\cos(m\sigma/\alpha_1)$ or $\cos(m(\sigma-\pi\alpha_1)/ \alpha_2)$).
Cut-off of the label  $p$ by $L$ is equivalent to discretizing 
the world sheet parameter
$\sigma$ to $L$ points.  Through the overlap 
given by the vertex, there exist
$L_1=\frac{\alpha_1}{\alpha_1+\alpha_2} L=-\beta L$ 
(resp. $L_2=\frac{\alpha_2}{\alpha_1+\alpha_2} L=(1+\beta) L$)
points on the first (resp. the second) closed string. 
While this reasoning may look weak, it turns out to be the unique choice which 
correctly produces the open string spectrum 
including higher modes. 

With this regularization, we obtain
\begin{equation}
 [\cdots]=\lim_{L\rightarrow \infty}
\left(\sum_{m=1}^{L_1}{1\over m}-\sum_{p=1}^{L}{1\over p}\right)=
\log (-\beta)\,.
\end{equation}
We have obtained a very compact result
for $E_1$,
\begin{eqnarray}
 E_1
&=&-a^{(+)\dagger}{\cal O}a^{(-)\dagger}
-\lambda^{-\beta(\sigma+\theta_1)+\pi}a^{\dagger}
+{\log (-\beta)\over 2} k_{\mu}G^{\mu\nu}k_{\nu}\,.\label{eq:E1}
\end{eqnarray}
We can derive $E_2$ similarly,
\begin{eqnarray}
 E_2=-a^{(+)\dagger}{\cal O}a^{(-)\dagger}-\lambda^{-(\beta+1)
(\pi-\sigma-\theta_2)}a^{\dagger}
+{\log (1+\beta)\over 2} k_{\mu}G^{\mu\nu}k_{\nu}\,.\label{eq:E2}
\end{eqnarray}
Coming back to Eqs.(\ref{eq:VBB}),(\ref{eq:BVB}),(\ref{eq:E1}),(\ref{eq:E2}),
\begin{eqnarray}
&&|\delta_T \Phi_B(\alpha_1)*\Phi_B(\alpha_2)\rangle
=c_B\,\oint {d\sigma\over 2\pi}
(-\beta)^{{1\over 2}k_{\mu}G^{\mu\nu}k_{\nu}}
\wp e^{ik_ix^i}
\oint {d\theta_1\over 2\pi}
e^{-a^{(+)\dagger}{\cal O}a^{(-)\dagger}-\lambda^{-\beta(\sigma+\theta_1)+\pi}
a^{\dagger}}\nonumber\\
&&~~~~~~~~~~~~~~~\times e^{c^{(+)\dagger}\bar{c}^{(-)\dagger}
+c^{(-)\dagger}\bar{c}^{(+)\dagger}}
\bar{c}_0|k_{\mu},x^i,\alpha_1+\alpha_2\rangle
\nonumber\\
&&=c_B\,(-\beta)^{{1\over 2}k_{\mu}G^{\mu\nu}k_{\nu}}
e^{ik_ix^i}
\oint {d\theta_3\over 2\pi}{d\sigma'\over 2\pi}
e^{-a^{(+)\dagger}{\cal O}a^{(-)\dagger}-\lambda^{-\beta\sigma'+\pi+\theta_3}
a^{\dagger}}e^{c^{(+)\dagger}\bar{c}^{(-)\dagger}
+c^{(-)\dagger}\bar{c}^{(+)\dagger}}
\bar{c}_0|k_{\mu},x^i,\alpha_1+\alpha_2\rangle
\nonumber\\
&&=(-\beta)^{{1\over 2}k_{\mu}G^{\mu\nu}k_{\nu}}c_B\,
|\delta_T \Phi_B(\alpha_1+\alpha_2)\rangle
\,,
\end{eqnarray}
and similarly
$
 |\Phi_B(\alpha_1)*\delta_T \Phi_B(\alpha_2)\rangle
=(1+\beta)^{{1\over 2}k_{\mu}G^{\mu\nu}k_{\nu}}c_B\,
|\delta_T\Phi_B(\alpha_1+\alpha_2)\rangle
$.
By putting these two equations into Eq.(\ref{eq:EOM}),
our  proof  of Eq.(\ref{eq_on-shell}) is finished:
 \begin{eqnarray}
 &&|\delta_T \Phi_B(\alpha_1)*\Phi_B(\alpha_2)\rangle 
 +|\Phi_B(\alpha_1)*\delta_T \Phi_B(\alpha_2)\rangle
 =c_B|\delta_T \Phi_B(\alpha_1+\alpha_2)\rangle\nonumber\\
 &&~~~~~~~~~~~~~~~\leftrightarrow~~~ (-\beta)^{\frac{1}{2}k_\mu G^{\mu\nu} k_\nu} +
 (1+\beta)^{\frac{1}{2} k_\mu G^{\mu\nu} k_\nu}=1\nonumber\\
 &&~~~~~~~~~~~~~~~\leftrightarrow~~~ k_{\mu}G^{\mu\nu}k_{\nu}=2\,.
 \end{eqnarray}
This is exactly  the on-shell condition 
of open string tachyon on the D$p$-brane.\footnote{
The on-shell condition of the perturbative {\it closed} tachyon is $p^2=8$
in our convention after \cite{HIKKO2}.
}

\subsection{Vector type fluctuation
\label{sec:vector}
}

Next, we consider a vector type fluctuation of the form of
Eq.(\ref{eq_var_vector}), which after using
the properties of the boundary state (\ref{eq:V_VB}), is equivalent to
\begin{eqnarray}
 |\delta_V \Phi_B(\alpha)\rangle=
\oint{d\sigma\over 2\pi }
(d^{\sigma}\cdot a^{\dagger})\,V_T(\sigma)|\Phi_B(\alpha)\rangle\,,
\label{eq:delVB}
\end{eqnarray}
where 
and  $d^{\sigma}$ is given by  $\zeta_{\mu}$ as
\begin{eqnarray}
 d^{\sigma}&=&
\zeta
\left(-
{1+{\cal O}^T\over 2}e^{-in\sigma},{1+{\cal O}\over 2}e^{in\sigma}
\right) C^{1\over 2}
=\zeta\left(i(\sin n\sigma)C^{1\over 2}(1,1){\cal P}_+
-(\cos n\sigma)C^{1\over 2}(1,-1){\cal P}_-
\right).
~~~~~\label{eq:dsigma}
\end{eqnarray}
We can compute the $*$ product of $\delta_V \Phi_B$ and $\Phi_B$
using the technique of Eq.(\ref{eq:del_lambda}):
\begin{eqnarray}
 &&|(\delta_V \Phi_B(\alpha_1))*\Phi_B(\alpha_2)\rangle
=\oint {d\sigma\over 2\pi}
\left(-d^{\sigma}{\partial\over \partial\lambda}|e^{-\lambda a^{\dagger}}
\Phi_B(\alpha_1)*\Phi_B(\alpha_2)\rangle\right)_{\lambda
=\lambda^{\sigma}}\nonumber\\
&&=c_B \wp \oint {d\sigma\over 2\pi}\oint {d\theta_1\over 2\pi}
(-\beta)^{{1\over 2}k_{\mu}G^{\mu\nu}k_{\nu}}\mathcal{D}_1\,
e^{-\lambda^{-\beta(\sigma+\theta_1)+\pi}a^{\dagger}}
|\Phi_B(\alpha_1+\alpha_2)\rangle\,,\label{eq:VB*B}\\
 &&|\Phi_B(\alpha_1)*(\delta_V \Phi_B(\alpha_2))\rangle
=\oint {d\sigma\over 2\pi}
\left(-d^{\sigma}{\partial\over \partial\lambda}
|\Phi_B(\alpha_1)*e^{-\lambda a^{\dagger}}\Phi_B(\alpha_2)
\rangle\right)_{\lambda=\lambda^{\sigma}}\nonumber\\
&&=c_B \wp \oint {d\sigma\over 2\pi}\oint {d\theta_2\over 2\pi}
(1+\beta)^{{1\over 2}k_{\mu}G^{\mu\nu}k_{\nu}}\mathcal{D}_2\,
e^{-\lambda^{-(\beta+1)(\pi-\sigma-\theta_2)}a^{\dagger}}
|\Phi_B(\alpha_1+\alpha_2)\rangle\,,\label{eq:B*VB}
\end{eqnarray}
where $\lambda^{\sigma}$ is given by Eq.(\ref{eq:lambda^sigma}).

There are three terms which contribute to $\mathcal{D}_1$ :
\begin{equation}
\mathcal{D}_1=d^{\sigma+\theta_1}(\cdots)a^{\dagger}
+d^{\sigma+\theta_1}(\cdots){\bf P}
+d^{\sigma+\theta_1}(\cdots)\lambda^{\sigma+\theta_1}\,.\label{eq:D1_1}
\end{equation}
The main contribution comes from the first term:
\begin{equation}
 \mathcal{D}_1=-\beta\, d^{-\beta(\sigma+\theta_1)+\pi}\cdot
a^{\dagger}+\cdots.
\end{equation}
We show the details of computation and other terms in appendix 
\ref{sec:vectordetail}.
Similarly, we obtain ${\cal D}_2$ in Eq.(\ref{eq:B*VB}) as
\begin{equation}
\mathcal{D}_2=(\beta+1) d^{-(\beta+1)(\pi-\sigma-\theta_2)}\cdot a^{\dagger}
+\cdots.
\end{equation}
Noting the integration over the interval $2\pi$ which is caused by
projection $\wp$,
the sum of Eqs.(\ref{eq:VB*B}) and (\ref{eq:B*VB})
becomes
\begin{eqnarray}
 &&|(\delta_V \Phi_B(\alpha_1))*\Phi_B(\alpha_2)\rangle 
+ |\Phi_B(\alpha_1)*(\delta_V \Phi_B(\alpha_2))\rangle
\nonumber\\
&=&
((-\beta)^{{1\over 2}k_{\mu}G^{\mu\nu}k_{\nu}+1}
+(1+\beta)^{{1\over 2}k_{\mu}G^{\mu\nu}k_{\nu}+1})
c_B|\delta_V\Phi_B(\alpha_1+\alpha_2)\rangle\nonumber\\
&&+((-\beta)^{{1\over 2}k_{\mu}G^{\mu\nu}k_{\nu}}
-(1+\beta)^{{1\over 2}k_{\mu}G^{\mu\nu}k_{\nu}})\left[
-i\zeta_\mu G^{\mu\nu}k_\nu \sum_{p=1}^\infty
\frac{\sin^2p\pi\beta}{\pi p}c_B|\delta_T \Phi_B(\alpha_1+\alpha_2)\rangle
+\cdots\right]
.\label{eq:-bb+1}
\end{eqnarray}
The remaining finite terms $[\cdots]$ are given in Eq.(\ref{eq:vector_LHS}).
{}From the first line we have obtained the on-shell condition
for vector type fluctuation (\ref{eq:delVB}):
\begin{eqnarray}
&&|(\delta_V \Phi_B(\alpha_1))*\Phi_B(\alpha_2)\rangle 
+ |\Phi_B(\alpha_1)*(\delta_V \Phi_B(\alpha_2))\rangle=
 c_B|\delta_V\Phi_B(\alpha_1+\alpha_2)\rangle\nonumber\\
&&~~~\leftrightarrow~~~~~~~~~
k_{\mu}G^{\mu\nu}k_{\nu}=0\,.
\label{eq_massless}
\end{eqnarray}
The interpretation of the second line is more subtle.
While the prefactor vanishes when (\ref{eq_massless}) is imposed,
$\zeta_\mu G^{\mu\nu}k_\nu$ 
also has a divergent coefficient $\sum_p \sin^2(p\pi \beta)/\pi p$.
In the regularization scheme we have used so far, we can not
make a definite statement whether this coefficient as a whole vanishes
or not. In this sense, it is not clear whether the
idempotency relation implies the transversality condition 
$\zeta_\mu G^{\mu\nu}k_\nu$=0.
We note that, as we have already commented, the vector type
deformation has the gauge symmetry (\ref{eq:gaugeinv})
which is the correct feature of the gauge particle.


It may be of some interest to compare it with
the analysis in VSFT \cite{HK}. 
While our variation $\delta_V\Phi_B$ has a unique form
of $d^{\sigma}$ (\ref{eq:dsigma}),  its counterpart
$d^{\mu}_n$ (Eq.(4.29) in Ref.\cite{HK}) of  VSFT
was arbitrary.  Actually they are all gauge degrees of freedom in VSFT
except for one \cite{Imamura}. As for the ordinary gauge
transformation, it was reproduced only 
after using regularization \cite{HKT}. While
we have discussed a close analogy of 
our analysis with VSFT, the gauge structure is very different.

The analysis of higher modes is more complicated
because of the treatment of the interaction point.
However, the leading term for the level $n$ perturbation
$\delta_n$
has the following simple structure,
\begin{eqnarray}
 |\delta_n \Phi_B*\Phi_B\rangle 
 + |\Phi_B*\delta_n\Phi_B\rangle&=&
 ((-\beta)^{{1\over 2}k_{\mu}G^{\mu\nu}k_{\nu}+n}
+ (1+\beta)^{{1\over 2}k_{\mu}G^{\mu\nu}k_{\nu}+n})
c_B|\delta_n \Phi_B\rangle+\cdots\,,
 \end{eqnarray}
 which gives the correct mass-shell condition
 for such vertices
$
{1\over 2}k_{\mu}G^{\mu\nu}k_{\nu}=1-n\,.
$
On the other hand, the cancellation
of the contributions from the interaction point
 will give a very nontrivial test of
our scenario that the idempotency condition of closed
string field would give the correct spectrum and symmetry
of the open string. 


\section{Discussion}
\label{sec:discussion}
We have seen that the ``vacuum version'' of closed
string field theory embodies the 
basic goals of the VSFT proposal;
namely it has a family of  exact solutions that correspond to 
various D-branes.
In our case, all of the basic types of the boundary states 
in the flat background (D$p$-brane with the flux)
appear as the exact solutions. 
Furthermore,  the infinitesimal variation of the solutions
produces the correct spectrum of the 
open string living on the D-brane (at least lower lying modes)
with the correct gauge symmetry.

What is the ``vacuum version'' of closed string field theory?
Resemblance of the action of HIKKO's string field theory
 \begin{equation}
  S=\frac{1}{2} \Phi\cdot Q\Phi +
\frac{1}{3} \Phi\cdot (\Phi*\Phi)
 \end{equation}
with Witten's open string field theory is one of the
encouraging points to suspect the existence of such a theory.
The computation of the tachyon vacuum
is parallel to the open string case \cite{G-R}
and will be possible at least numerically.
As in the VSFT proposal, one may conjecture that the equation of
motion at this ``vacuum'' may be written as
Eq.(\ref{eq:B2=B}).  We have observed 
the close analogies of the structure of the pure ghost kinetic term 
at the end of section \ref{sec:Gaussian}.
At this vacuum, there would be no propagating degree of
freedom both in the closed and open string sectors.
There exist, however, the nonperturbative solutions
-- the boundary states.

{}It is tempting to conjecture that, as in the VSFT proposal,
the re-expansion of the theory around the solution
produces open string field theory.  
By the assumption of the vacuum theory,
 there is no closed string propagation at the tree level.
On the other hand, the open string becomes physical.
The BRST charge at the new vacuum would be
\begin{equation}
{\cal Q}|\Phi\rangle  = \hat{\alpha}^2 \frac{\partial}{\partial \bar{c}_0}
 |\Phi\rangle -2 | \hat{\Phi}_B * \Phi\rangle\,,\quad
\hat{\Phi}_B = \lim_{\alpha\rightarrow 0/\infty}
\frac{\alpha^2}{{\frak c}_{\beta=-1/2}V_{d-p-1}}\Phi_B(\alpha)\,,
\end{equation}
which is formally nilpotent by the Jacobi identity.
$\hat{\alpha}^2$ factor in front of $\bar c_0$ derivative
is needed to make this part a derivation.
Here we need to take the limit $\alpha\rightarrow 0
\mbox{ or }\infty$ since $\alpha$ parameter is preserved
by the star product.  More detailed
examination of this scenario will be presented 
in the future study.

The use of the closed string degree of freedom has
a definite advantage in describing the physical process
involving the D-brane, for example, in the time-dependent solutions
of the D-brane decay. In such a situation, the r\^ole of
the closed strings seems more important than the open strings
\cite{ref:decay}.
If we use the open string fields alone, the treatment of
closed strings becomes singular while in our approach
it is encoded as the fundamental degrees of freedom.
Of course, to proceed in this direction, we need to understand
how the propagating degrees of freedom appear in the closed
string sector, which would be the most important issue
in our proposal.

We describe our scenario as
a possible physical interpretation of the idempotency equation of
the boundary states.  We  do not deny the other possibilities
at this point.  Since Eq.(\ref{eq:B2=B})
is a mathematically rigorous 
statement, it will play a fundamental r\^ole even
if our scenario might not be so accurate.

Finally we have obtained
the boundary states which satisfy the idempotency relation,
Eq.(\ref{eq:B2=B}), rather than the equation of
motion of full string field theory,
\begin{equation}
  Q\Phi + \Phi*\Phi =0\,\,.
\end{equation}
For example, in \cite{DiVecchia}, it was argued that 
the asymptotic behavior of $\frac{1}{L_0+\bar{L}_0}|B\rangle$
coincides with the supergravity solution at the linear level.
However, at the nonlinear level, it is easy to see that
it fails to be the solution of HIKKO's equation of motion.
The settlement of this apparent conflict is another good
challenge in the future.

\begin{center}
\noindent{\large \textbf{Acknowledgments}}
\end{center}

We would like to thank K.~Ohmori for valuable discussions and comments.
I.K. would like to thank H.~Hata, S.~Imai, T.~Kawano and H.~Kogetsu
for useful conversation.
I.K. is supported in part by JSPS Research Fellowships
for Young Scientists. Y.M. is supported in part by Grant-in-Aid (\#
13640267) from the Ministry of Education, Science, Sports and Culture of
Japan.

\appendix

\section{Notations and conventions
\label{sec:conventions}
}
We give a summary of our convention of the oscillators
and the vacuum used in the text.
The mode expansions of basic oscillators are given as follows
\cite{HIKKO2}:
\begin{eqnarray}
&& X^M(\sigma)=\frac{1}{\sqrt{\pi}}\left\{
x^M+{i\over 2}\sum_{n\ne 0}{1\over n}\left(\alpha_n^{(+)M}
-\alpha_{-n}^{(-)M}\right)e^{i n\sigma}\right\}\,,\,\nonumber\\
&& P_M(\sigma)={1\over 2\sqrt{\pi}}
\left\{p_M+\eta_{MN}\sum_{n\ne 0}\left(\alpha_n^{(+)N}+
\alpha_{-n}^{(-)N}\right)e^{in\sigma}\right\}\,,\nonumber\\
&& \bar{c}(\sigma)={1\over 2\sqrt{\pi}}\left\{
\bar{c}_0+\sum_{n\ne 0}\left(\bar{c}_n^{(+)}+\bar{c}_{-n}^{(-)}\right)
e^{in\sigma}
\right\}\,,\quad
i\pi_{\bar{c}}(\sigma)={1\over 2\sqrt{\pi}}\left\{
2{\partial\over \partial \bar{c}_0}+\sum_{n\ne 0}
\left(c_n^{(+)}+c_{-n}^{(-)}\right)e^{in\sigma}
\right\}\,,\\
&& c(\sigma)=-{1\over 2\sqrt{\pi}}\left\{
i{\partial\over \partial \pi_{\bar{c}}^0}+
\sum_{n\ne 0}\left(c_n^{(+)}-c_{-n}^{(-)}\right)e^{in\sigma}
\right\}\,,\,
i\pi_c(\sigma)=-{1\over 2\sqrt{\pi}}\left\{
-2i\pi_c^0+\sum_{n\ne 0}\left(\bar{c}_n^{(+)}-
\bar{c}_{-n}^{(-)}\right)e^{in\sigma}\right\}\,.\nonumber
\end{eqnarray}
We note that 
$\bar{c}(\sigma)$ is written more often 
as $b(\sigma)$ in the literature.

Commutation relations of nonzero modes are
\begin{eqnarray}
&&\left[\alpha_m^{(\pm)M},\alpha_n^{(\pm)N}\right]=m\delta_{m+n,0}\eta^{MN}\,,
~~~
\left\{c_m^{(\pm)},\bar{c}_n^{(\pm)}\right\}=\delta_{m+n,0}\,.
\end{eqnarray}
We often use the notation,
\begin{eqnarray}
\alpha_n^{(\pm)M}=\sqrt{n}a^{(\pm)M}_n\,,
\alpha_{-n}^{(\pm)M}=\sqrt{|n|}a^{(\pm)M\dagger}_n\,,~~
c^{(\pm)\dagger}_m=c^{(\pm)}_{-m}\,,
~~\bar{c}^{(\pm)\dagger}_m=\bar{c}^{(\pm)}_{-m}\,,~~~
(m,n>0)
\end{eqnarray}
which satisfy
\begin{equation}
 [a_m^{(\pm)M},a_n^{(\pm)N\dagger}]=\delta_{m,n}\eta^{MN}\,,~
~~~\left\{c_m^{(\pm)},\bar{c}_n^{(\pm)\dagger}\right\}=
\left\{c_m^{(\pm)\dagger},\bar{c}_n^{(\pm)}\right\}=\delta_{m,n}\,.
\end{equation}
Matter zero modes are represented by $\hat{a}_0,\hat{a}_0^{\dagger}$ as
\begin{equation}
\hat{x}^M={\frac{i}{2}}(\hat{a}_{0}^M-\hat{a}_{0}^{M\dagger })\,,\quad
\hat{p}_{M}=\eta _{MN}(\hat{a}_{0}^{N}+\hat{a}_{0}^{N\dagger })\,,\quad 
[\hat{x}^M,\hat{p}_{N}]=i\delta_N^M\,,
\quad [\hat{a}_{0}^M,\hat{a}_{0}^{N\dagger}]=\eta ^{MN}\,,
\end{equation}
and their eigenstates are given by
\begin{eqnarray}
&&\langle x| =\langle 0|e^{{\frac{1}{2}}\hat{a}_{0}^{2}+2ix\hat{a}_{0}
-x^{2}}\left(2/\pi\right)^{\frac{d}{4}}\,,~~~~
|x\rangle =\left(2/\pi\right)^{\frac{d}{4}}
e^{{\frac{1}{2}}\hat{a}_{0}^{\dagger 2}
-2ix\hat{a}_{0}^{\dagger }-x^{2}}|0\rangle \,,\\
&&\langle x|\hat{x}^M=\langle x|x^M\,,~~~
\hat{x}^M|x\rangle =x^M|x\rangle \,,~~~~~
\langle x|x'\rangle =\delta^d(x-x')\,,\\
&&\langle p|=\langle 0|e^{-{\frac{1}{2}}\hat{a}_{0}^{2}+\hat{a}_{0}p
-\frac{1}{4}p^{2}}\left( 2\pi\right)^{\frac{d}{4}}\,, ~~~
|p\rangle =\left(2\pi\right)^{\frac{d}{4}}
e^{-{\frac{1}{2}}\hat{a}_{0}^{\dagger 2}
+\hat{a}_{0}^{\dagger}p-{\frac{1}{4}}p^{2}}|0\rangle \,,   \\
&&\langle p|\hat{p}^M=\langle p|p^M\,,~~~
\hat{p}^M|p\rangle=p^M|p\rangle\,,~~~~~
\langle p|p^{\prime }\rangle 
=(2\pi)^{d}\delta^{d}(p-p^{\prime })\,,   \\
&&\langle p|x\rangle =e^{-ipx},\quad 
\langle x|p\rangle =e^{ipx}\,,\quad
 \langle 0|\hat{a}_0^{\dagger}=0\,,~~~\hat{a}_0|0\rangle=0\,.
\end{eqnarray}
Similarly $\alpha$-dependent part is treated as an analogue of
matter zero modes,
\begin{equation}
\hat{\alpha}|\alpha\rangle=\alpha|\alpha\rangle\,,\quad
\langle \alpha|\hat{\alpha}=\langle \alpha|\alpha\,,\quad
\langle \alpha|\alpha'\rangle=(2\pi)\delta(\alpha-\alpha')\,.
\end{equation}
On the ghost zero mode, the bra-ket convention is
\begin{equation}
{\partial \over \partial \bar{c}_0}|0\rangle=0\,,\quad
\langle 0|{\overleftarrow{\partial} \over \partial \bar{c}_0}=0\,.
\end{equation}

\section{Gaussian formulae
\label{sec:Gaussian formula}
}

In string field theories using oscillator representation,
 we often encounter computations of the form
$e^{aMa}e^{a^{\dagger}Na^{\dagger}}|0\rangle$.
We show useful formulae for this type computations.
These are proved by inserting coherent states and performing Gaussian
 integration.

For the matter sector with bosonic oscillators
\begin{equation}
 [a_m,a_n^\dagger]=\delta_{mn},\ ~~~~~a_n|0\rangle=0,~~n\ge 1\,,
\end{equation}
we have
\begin{eqnarray}
&&\exp\left({1\over2}aMa+\lambda a\right)\exp\left({1\over2}a^\dagger 
Na^\dagger+\mu a^\dagger\right)|0\rangle\nonumber \\
&&={1\over\sqrt{{\rm det}(1-MN)}}\exp\left({1\over2}\lambda N(1-MN)^{-1}
\lambda+{1\over2}\mu M(1-NM)^{-1}\mu+\lambda(1-NM)^{-1}\mu\right)\nonumber \\
&&\cdot \exp\left((\lambda N+\mu)(1-MN)^{-1}a^\dagger+{1\over2}a^\dagger 
N(1-MN)^{-1}a^\dagger\right)|0\rangle \,,
\label{eq:gaussian_matter}
\end{eqnarray}
where $M,N$ are symmetric matrices.

For the ghost sector with fermionic oscillators
\begin{equation}
 \{c_n,b_m\}=\delta_{n+m,0}\,,~~c_n|+\rangle=0\,,~~n\ge 0\,,~~
 b_n|+\rangle=0\,,~~n\ge 1\,,
~~~~~~c^{\dagger}_n:=c_{-n},\quad b^{\dagger}_n:=b_{-n}\,,~~n\ge 1\,,
\end{equation}
we have
\begin{eqnarray}
&&\exp(cAb+c_0\alpha b+c\mu+\nu b+c_0\gamma)\exp(c^\dagger Bb^\dagger
+c^\dagger\beta b_0+c^\dagger\rho+\sigma b^\dagger+\delta b_0)|+\rangle
\nonumber\\
&=&\det(1+BA)\det \Delta \cdot e^{E_1+E_0} |+\rangle \,,
\label{eq:gaussian_ghost}
\end{eqnarray}
where
\begin{eqnarray}
&&\Delta=1+\alpha(1+BA)^{-1}\beta,\nonumber \\
&&E_1=c^\dagger(1+BA)^{-1}Bb^\dagger+c^\dagger(1+BA)^{-1}(\rho-B\mu)
+(\nu B+\sigma)(1+AB)^{-1}b^\dagger\nonumber \\
&&+\nu(1+BA)^{-1}(\rho-B\mu)-\sigma(1+AB)^{-1}(A\rho+\mu),\nonumber \\
&&E_0=-c^\dagger(1+BA)^{-1}\beta\Delta^{-1}(\alpha(1+BA)^{-1}Bb^\dagger-b_0)
-c^\dagger(1+BA)^{-1}\beta\Delta^{-1}(\alpha(1+BA)^{-1}(\rho-B\mu)+\gamma)
\nonumber \\
&&-((\nu-\sigma A)(1+BA)^{-1}\beta+\delta)\Delta^{-1}(\alpha(1+BA)^{-1}
Bb^\dagger-b_0) \nonumber \\
&&-((\nu-\sigma A)(1+BA)^{-1}\beta+\delta)\Delta^{-1}(\alpha(1+BA)^{-1}
(\rho-B\mu)+\gamma)\,.
\end{eqnarray}
In particular, if there are no terms dependent on zero mode, the above formula
is simplified as $\Delta=1\,,~E_0=0$.
In the computation of Eq.(\ref{eq:Gaussian_star}), we use it
for $\alpha=\mu=\nu=\gamma=0$ case.

\section{Relations among Neumann coefficients of light-cone type SFT
\label{sec:formulae}
}

\subsection{Definitions of Neumann coefficients}
Neumann coefficients are used to define 3-string vertex
 $|V(1,2,3)\rangle$ 
which represents connection conditions of string world sheets and 
encodes string interactions.

In Eq.(\ref{eq:v123}), we used light-cone type Neumann coefficients
$\bar{N}_{mn}^{rs},\bar{N}_m^r$
which
are explicitly given by 
\cite{GS}\cite{HIKKO1}\cite{HIKKO2}:
\begin{eqnarray}
 \bar{N}_{mn}^{rs}&=&-\alpha_1\alpha_2\alpha_3
\left({\alpha_r\over m}+{\alpha_s\over n}\right)^{-1}
\bar{N}^r_m\bar{N}^s_n\,,\label{eq_Neum2}\\
\bar{N}^r_m&=&{1\over \alpha_r}f_m\left(-\alpha_{r+1}/\alpha_r\right)
e^{m\tau_0/\alpha_r}\,,~~~~(\alpha_4:=\alpha_1)\,
\label{eq_Neum3}\\
f_n(x)&=&{\Gamma(nx)\over n!\Gamma(nx-n+1)}\,.\label{eq_Neum4}
\end{eqnarray}
We also use the notation
\begin{equation}
\tilde{N}_{mn}^{rs}:=\sqrt{m}\bar{N}_{mn}^{rs}\sqrt{n}\,,~~~~~
\tilde{N}_m^r:=\sqrt{m}\bar{N}_m^r\,.
\end{equation}
They satisfy relations \cite{Yoneya}
\begin{eqnarray}
&&\sum_{t=1}^3\sum_{p=1}^{\infty}\tilde{N}^{rt}_{mp}\tilde{N}^{ts}_{pn}
=\delta_{r,s}\delta_{m,n}\,,~~~~
\sum_{t=1}^3\sum_{p=1}^{\infty}\tilde{N}^{rt}_{mp}\tilde{N}^{t}_{p}
=-\tilde{N}^r_m\,,~~~~
\sum_{t=1}^3\sum_{p=1}^{\infty}\tilde{N}^t_p\tilde{N}^t_p
={2\tau_0\over \alpha_1\alpha_2\alpha_3}\,.
\label{eq:yoneya}
\end{eqnarray}
It is convenient to rewrite them using matrix representations as
\begin{eqnarray}
\tilde{N}^{rs}_{mn}&=&(C^{1\over 2}\bar{N}^{rs}C^{1\over 2})_{mn}=
\delta_{m,n}\delta_{r,s}-2(A^{(r)T}\Gamma^{-1}A^{(s)})_{mn}\,,\\
\tilde{N}^r_m&=&(C^{1\over 2}\bar{N}^r)_m=
-(A^{(r)T}\Gamma^{-1}B)_m\,,
\end{eqnarray}
where, for $|\alpha_1|+|\alpha_2|=|\alpha_3|$, they are given by
\begin{eqnarray}
  A_{mn}^{~(1)}&=&2\sqrt{n\over m}(-1)^m{1\over \pi\alpha_1}
\int^{\pi\alpha_1}_0d\sigma\cos{n\sigma\over\alpha_1}
\cos{m\sigma\over\alpha_3}
=-{2\over\pi}\sqrt{mn}(-1)^{m+n}{\beta\sin m\pi\beta\over n^2-m^2\beta^2}\,,\\
A_{mn}^{~(2)}&=&2\sqrt{n\over m}(-1)^m{1\over \pi\alpha_2}
\int^{\pi(\alpha_1+\alpha_2)}_{\pi\alpha_1}d\sigma
\cos{n(\sigma-\pi\alpha_1)\over\alpha_2}\cos{m\sigma\over\alpha_3}\\
&=&-{2\over\pi}\sqrt{mn}(-1)^m
{(\beta + 1)\sin m\pi\beta\over n^2-m^2(\beta +1)^2}\,,\\
A_{mn}^{(3)}&=&\delta_{m,n}\,,\\
\Gamma_{mn}&=&\delta_{m,n}+\sum_{r=1}^2(A^{(r)}A^{(r)T})_{mn}
=\Gamma_{nm}\,,\\
B_m&=&-{2\over\pi}{\alpha_3\over\alpha_1\alpha_2}m^{-{3\over 2}}(-1)^m
\sin m\pi\beta\,,\\
C_{mn}&=&m\delta_{m,n}\,.
\end{eqnarray}
Here, we used the notation
\begin{equation}
 \beta={\alpha_1\over \alpha_3}\,,~~\beta+1=-{\alpha_2\over \alpha_3}\,,
~~~{\rm for}~~~\alpha_1+\alpha_2+\alpha_3=0\,.
\end{equation}
We note that $-1<\beta<0$ in this case.

\subsection{Relations among overlap coefficients $A$, $B$}
We list some relations between the
coefficients $A$  and $B$ which were proved in \cite{GS}:
\begin{eqnarray}
&&-{\alpha_r\over \alpha_3}(C^{-1}A^{(r)T}CA^{(s)})_{mn}
=\delta_{r,s}\delta_{m,n}\,,~~~(r,s=1,2)\label{eq:Nise_inv}\\
&&(A^{(r)T}CB)_m=0\,,~~~~~~(r=1,2)\label{eq:A_zeromode}\\
&&{1\over 2}\alpha_1\alpha_2 B^TCB=1\,,\\
&&(\Gamma^{-1}C^{-1}A^{(r)})_{mn}=
(C^{-1}A^{(r)})_{mn}+{\alpha_r\over \alpha_3}(\Gamma^{-1}A^{(r)}C^{-1})_{mn}
\,,~~~(r=1,2)\\
&&{\alpha_3\over m}\delta_{m,n}
+\sum_{r=1}^2\alpha_r(A^{(r)}C^{-1}A^{(r)T})_{mn}=
{1\over 2}\alpha_1\alpha_2\alpha_3B_mB_n\,,\\
&&(\Gamma C^{-1}\Gamma)_{mn}=(C^{-1}\Gamma)_{mn}+(\Gamma C^{-1})_{mn}
-{1\over 2}\alpha_1\alpha_2B_mB_n\,,\\
&&(C^{-1})_{mn}-(C^{-1}\Gamma^{-1})_{mn}-(\Gamma^{-1}C^{-1})_{mn}
+{1\over 2}\alpha_1\alpha_2(\Gamma^{-1}B)_m(\Gamma^{-1}B)_n=0\,,\\
&&{1\over 2}\alpha_1\alpha_2\alpha_3(A^{(r)T}\Gamma^{-1}B)_m
(A^{(s)T}\Gamma^{-1}B)_n
=-\alpha_r(C^{-1})_{mn}\delta_{r,s}+\left({\alpha_r\over m}
+{\alpha_s\over n}\right)(A^{(r)T}\Gamma^{-1}A^{(s)})_{mn}\nonumber\\
&&~~~~~~~~(r,s=1,2,3)\\
&&\left((1-A^T\Gamma^{-1} A)^{-1}\right)^{rs}_{mn}
=\delta_{r,s}\delta_{m,n}+(A^{(r)T}A^{(s)})_{mn}\,,~~~~~
(r,s=1,2)\\
&&\left((A^T\Gamma^{-1}A)^{-1}\right)^{rs}_{mn}=
\delta_{r,s}\delta_{m,n}+{\alpha_3^2\over \alpha_r\alpha_s}
(CA^{(r)T}C^{-2}A^{(s)}C)\,,~~~~~~(r,s=1,2)\\
&&B^T\Gamma^{-1}B={2\tau_0\over \alpha_1\alpha_2\alpha_3}\,.
\end{eqnarray}
In particular, the infinite matrices $(A_{mn}^{~(1)},A_{mn}^{~(2)})$
is invertible.\footnote{
One might think $-{\alpha_r\over \alpha_3}(C^{-1}A^{(r)T}C)_{mn}$ is also
an ``inverse'' from Eq.(\ref{eq:Nise_inv}). However, this matrix has zero
mode (\ref{eq:A_zeromode}).
This kind of subtlety was noticed in Ref.\cite{BM1} 
for Witten's open string field theory.
}
Namely, we can find an inverse matrix:
\begin{equation} 
D^{(r)}_{mn}=-{\alpha_3\over \alpha_r}(CA^{(r)T}C^{-1})_{mn}\,.
\label{eq:Ddef}
\end{equation}
In fact, we can prove
\begin{eqnarray}
&&\sum_{k=1}^{\infty}D^{(r)}_{mk}A^{~(s)}_{kn}=\delta_{m,n}\delta_{r,s}\,,
~~~(r,s=1,2)
~~~~~~~\sum_{r=1}^2\sum_{k=1}^{\infty}
A_{mk}^{~(r)}D_{kn}^{(r)}=\delta_{m,n}\,,~~~
\label{eq:DAAD}
\end{eqnarray}
directly.
These relations are mainly based on the Fourier expansion \cite{GS}
\begin{eqnarray}
 &&\sum_{n=-\infty}^{\infty}{(-1)^ne^{iny}\over n+\alpha}
={\pi\over \sin \pi\alpha}e^{-i\alpha y}\,,~~~(-\pi<y<\pi)
\end{eqnarray}
or
\begin{eqnarray}
  &&{1\over \alpha}-2\alpha\sum_{n=1}^{\infty}
{(-1)^n\cos ny\over n^2-\alpha^2}
={\pi\cos\alpha y\over \sin \pi\alpha}\,,~~~(-\pi<y<\pi)
\label{eq:sum_cos}
\\
&&\sum_{n=1}^{\infty}(-1)^n{2n\sin ny\over n^2-\alpha^2}=
-{\pi\sin\alpha y\over \sin\pi \alpha}\,,~~~(-\pi<y<\pi)\,.
\label{eq:sum_sin}
\end{eqnarray}

\subsection{List of useful formulae related to matrix $n$ and
  $\tilde{N}^{r3}$
\label{sec:list-neumann}
}
We collect useful formulae associated with the Neumann
coefficients $n$ and $\tilde N^{r3}$ (see Eq.(\ref{eq_def_n})):
\begin{eqnarray}
((1-n^2)^{-1})^{(rs)} & = & \frac{1}{4}D^{(r)}\Gamma^2 D^{(s)T}
 =\frac{1}{4}(D^{(r)}D^{(s)T}+A^{(r)T}A^{(s)}+2\delta^{(rs)})\,,\\
((1-n)^{-1})^{(rs)} & = & \frac{1}{2} (\delta^{(rs)}+D^{(r)}D^{(s)T})\,,
\\
((1+n)^{-1})^{(rs)} & = & \frac{1}{2} (\delta^{(rs)}+A^{(r)T}A^{(s)})\,,
\\
(n(1-n)^{-1})^{(rs)} & = & \frac{1}{2}(D^{(r)}D^{(s)T}-\delta^{(rs)})\,,
\\
(n(1+n)^{-1})^{(rs)} & = & -\frac{1}{2}(A^{(r)T}A^{(s)}-\delta^{(rs)})\,,
\\
\sum_s ((1-n)^{-1})^{(rs)}\tilde{N}^{(s3)} & = & -D^{(r)}\,,\\
\sum_s ((1+n)^{-1})^{(rs)}\tilde{N}^{(s3)} & = &  - A^{(r)T}\,.
\end{eqnarray}

\subsection{Some formulae associated with $(\cos n\sigma)$,
$(\sin n\sigma)$ and $A^{(r)}$, $D^{(r)}$}
We list some more formulae which we use in computations
in \S \ref{sec:fluctuation}.

For the interval  $-\pi< \sigma < \pi$,
\begin{eqnarray}
&&\left[(\cos n\sigma)C^{-{1\over 2}}A^{(1)T}
\right]_m=
{1\over 2}\alpha_2B_m+{(-1)^m\over \sqrt{m}}\cos m\beta\sigma\,,\\
&&\left[(\sin n\sigma)C^{-{1\over 2}}
D^{(1)}
\right]_m=
-{(-1)^m\over \sqrt{m}}\sin m\beta\sigma\,,\\
 &&\left[(\sin n\sigma)C^{1\over 2}A^{(1)T}\right]_m=
\beta\sqrt{m}(-1)^m\sin m\beta\sigma\,,\\
 &&\left[(\cos n\sigma) C^{1\over 2}
D^{(1)}
\right]_m
=-\beta\sqrt{m}(-1)^m\cos m\beta\sigma
+\left(\sum_{n=1}^{\infty}(-1)^n\cos n\sigma +{1\over 2}\right)
{(-1)^m 2\over \sqrt{m}\pi}\sin m\pi\beta\nonumber\\
&&=-\beta\sqrt{m}(-1)^m\cos m\beta\sigma
-\left(\sum_{n=1}^{\infty}(-1)^n\cos n\sigma +{1\over 2}\right)
\alpha_2\beta B_m m\,. \label{eq:cosD}
\end{eqnarray}

For the interval  $-\pi< \pi - \sigma < \pi$,
\begin{eqnarray}
 &&\left[(\cos n\sigma)C^{-{1\over 2}}A^{(2)T}
\right]_m=
-{1\over 2}\alpha_1B_m+{1\over \sqrt{m}}\cos( m(\beta+1)(\pi-\sigma))\,,\\
&&\left[(\sin n\sigma)C^{-{1\over 2}}
D^{(2)}
\right]_m=
-{1\over \sqrt{m}}\sin (m(\beta+1)(\pi-\sigma))\,,\\
 &&\left[(\sin n\sigma)C^{1\over 2}A^{(2)T}\right]_m=
-(\beta+1)\sqrt{m}\sin (m(\beta+1)(\pi -\sigma))\,,\\
&&\left[(\cos n\sigma) C^{1\over 2}
D^{(2)}
\right]_m\nonumber\\
&&~~~~=(\beta+1)\sqrt{m}\cos (m(\beta+1)(\pi-\sigma))
-\left(\sum_{n=1}^{\infty}(-1)^n\cos n(\pi-\sigma) +{1\over 2}\right)
{(-1)^m 2\over \sqrt{m}\pi}\sin m\pi\beta\nonumber\\
&&~~~~=(\beta+1)\sqrt{m}\cos (m(\beta+1)(\pi-\sigma))
-\left(\sum_{n=1}^{\infty}(-1)^n\cos n(\pi-\sigma) +{1\over 2}\right)
\alpha_1(\beta+1)B_m m\,.~~~\label{eq:cosD2}
\end{eqnarray}

\section{Oscillators on the boundary state
\label{sec:Oscil Boundary}
}

The conditions, Eqs.(\ref{eq:BFi}),(\ref{eq:BFmu}),(\ref{eq:BFpi}),
of the boundary state $|B(F)\rangle$ corresponding to the D$p$-brane
can be rewritten in terms of the oscillators as follows.

\noindent Nonzero modes:
\begin{eqnarray}
&&\left(\alpha_n^{(+)i}-\alpha_{-n}^{(-)i}\right)|B(F)\rangle=0\,,
\label{eq_b_a1}\\
&&\left(\alpha_n^{(+)\mu}+{\cal O}^{\mu}_{~\nu}\alpha_{-n}^{(-)\nu}\right)
|B(F)\rangle=\left(\alpha^{(-)\mu}_{n}
+({\cal O}^T)^{\mu}_{~\nu}\alpha^{(+)\nu}_{-n}\right)|B(F)\rangle=0\,,
\label{eq_b_a2}\\
&&\left(c^{(+)}_n+c^{(-)}_{-n}\right)|B(F)\rangle=
\left(\bar{c}_n^{(+)}-\bar{c}_{-n}^{(-)}\right)|B(F)\rangle=0\,.
\label{eq_b_c}
\end{eqnarray}

\noindent Zero mode:
\begin{eqnarray}
 &&(\hat{x}^i-x^i)|B(F)\rangle=\hat{p}^{\mu}|B(F)\rangle=0\,,~~~~
{\partial\over \partial \bar{c}_0}|B(F)\rangle=0\,.
\end{eqnarray}

It is convenient to define new oscillators
$\alpha'_n$ on the boundary state $|B(F)\rangle$ 
to consider vertex operators on it:\footnote{
Here we defined $\epsilon(n)=\left\{
\begin{array}[tb]{cc}
 +1 & (n>0) \\
-1 & (n<0)
\end{array}
\right.\,.$
}
\begin{eqnarray}
 &&\alpha'{}^{(+)\mu}_n:={1\over \sqrt{2}}\left(\alpha_n^{(+)\mu}+
\epsilon(n){\cal O}^{\mu}_{~\nu}\alpha_{-n}^{(-)\nu}\right)\,,~~~~
\alpha'{}^{(-)\mu}_n:={1\over \sqrt{2}}\left(\alpha_n^{(-)\mu}+
\epsilon(n)({\cal O}^T)^{\mu}_{~\nu}\alpha_{-n}^{(+)\nu}\right)\,,\\
 &&\alpha'{}^{(+)i}_n:={1\over \sqrt{2}}\left(\alpha_n^{(+)\mu}-
\epsilon(n)\alpha_{-n}^{(-)i}\right)\,,~~~~
\alpha'{}^{(-)i}_n:={1\over \sqrt{2}}\left(\alpha_n^{(-)i}-
\epsilon(n)\alpha_{-n}^{(+)i}\right)\,,\\
&&\left[\alpha'{}^{(\pm)M}_m,\alpha'{}^{(\pm)N}_n\right]
=m\delta_{m+n,0}\eta^{MN}\,,~~~~
\alpha'{}^{(\pm)M}_n|B(F)\rangle=0\,,~~~~n\ge 1\,.
\end{eqnarray}
In terms of $\alpha'_n$ we can rewrite $X^M(\sigma),P^M(\sigma)$ as
\begin{eqnarray}
 X^{\mu}(\sigma)&=&{1\over \sqrt{\pi}}\left[
x^{\mu}+{i\over \sqrt{2}}\sum_{n\ne 0}{1\over n}\left(
\left({1-\epsilon(n){\cal O}^T\over 2}\right)^{\mu}_{~\nu}
\alpha'{}^{(+)\nu}_{n}
-\left({1+\epsilon(n){\cal O}\over 2}\right)^{\mu}_{~\nu}
\alpha'{}^{(-)\nu}_{-n}
\right)e^{in\sigma}
\right]
\,,\\
X^i(\sigma)&=&{1\over \sqrt{\pi}}\left[
x^i+{i\over \sqrt{2}}\sum_{n=1}^{\infty}{1\over n}\left(
\alpha'{}^{(+)i}_ne^{in\sigma}+\alpha'{}^{(-)i}_n e^{-in\sigma}
\right)
\right]
\,,\\
P^{\mu}(\sigma)&=&{1\over 2\sqrt{\pi}}\left[
p^{\mu}+\sqrt{2}\sum_{n\ne 0}\left(
\left({1+\epsilon(n){\cal O}^T\over 2}\right)^{\mu}_{~\nu}
\alpha'{}^{(+)\nu}_n
+
\left({1-\epsilon(n){\cal O}\over 2}\right)^{\mu}_{~\nu}
\alpha'{}^{(+)\nu}_{-n}
\right)e^{in\sigma}
\right]\,,\\
P^i(\sigma)&=&{1\over 2\sqrt{\pi}}\left[
p^i+\sqrt{2}\sum_{n=1}^{\infty}\left(\alpha'{}_{-n}^{(+)i}e^{-in\sigma}
+
\alpha'{}_{-n}^{(-)i}e^{in\sigma}
\right)
\right]\,.
\end{eqnarray}

We can define the normal ordering of tachyon vertex with respect to the new
oscillators\newline
$a'{}^{(\pm)}_n:=\alpha'{}^{(\pm)}_n/\sqrt{n},~
a'{}^{(\pm)\dagger}_n:=\alpha'{}^{(\pm)}_{-n}/ \sqrt{n},~~(n\ge 1)$
as
\begin{eqnarray}
V_T(\sigma)&=&{\cal N}
:e^{ik_M\sqrt{\pi} X^M(\sigma)}:\nonumber\\
&=&{\cal N}\exp\left(k_{\mu}\sum_{n=1}^{\infty}{1\over \sqrt{2n}}\left(
{1+{\cal O}^T\over 2}a'{}^{(+)\dagger}_{n}e^{-in\sigma}
+{1+{\cal O}\over 2}a'{}^{(-)\dagger}_{n}e^{in\sigma}
\right)^{\mu}\right)e^{ik_M\hat{x}^M}\nonumber\\
&&\times \exp\left(-k_{\mu}\sum_{n=1}^{\infty}{1\over \sqrt{2n}}
\left({1-{\cal O}^T\over 2}a'{}^{(+)}_ne^{in\sigma}
+{1-{\cal O}\over 2}a'{}^{(-)}_ne^{-in\sigma}\right)^{\mu}
\right)\nonumber\\
&&\times \exp\left(
-k_i\sum_{n=1}^{\infty}{1\over \sqrt{2n}}
\left(a'{}^{(+)}_ne^{in\sigma}
+a'{}^{(-)}_ne^{-in\sigma}
\right)^i\right)\,.
\end{eqnarray}
Then we have
\begin{eqnarray}
 V_T(\sigma)|B(F)\rangle&=&{\cal N}
\exp\left(k_{\mu}\sum_{n=1}^{\infty}{1\over \sqrt{2n}}\left(
{1+{\cal O}^T\over 2}a'{}^{(+)\dagger}_{n}e^{-in\sigma}
+{1+{\cal O}\over 2}a'{}^{(-)\dagger}_{n}e^{in\sigma}
\right)^{\mu}\right)e^{ik_M\hat{x}^M}|B(F)\rangle\nonumber\\
&=&
\exp\left(k_{\mu}\sum_{n=1}^{\infty}{1\over \sqrt{n}}\left(
{1+{\cal O}^T\over 2}a{}^{(+)\dagger}_{n}e^{-in\sigma}
+{1+{\cal O}\over 2}a{}^{(-)\dagger}_{n}e^{in\sigma}
\right)^{\mu}\right)e^{ik_M\hat{x}^M}|B(F)\rangle\,.~~~~~~~
\label{eq:V_TB}
\end{eqnarray}
In the last equation we rewrote again in terms of 
 original oscillators.\footnote{
Here we have chosen the normalization constant as
${\cal N}=e^{{1\over 2}k{1+{\cal O}\over 2}k\sum_{n=1}^{\infty}{1\over n}}$.
}

Similarly, we can consider the vector vertex on the boundary state as follows
\begin{eqnarray}
 V_V(\sigma)|B(F)\rangle
&=&{\cal N}:\zeta_M \sqrt{\pi} \partial_{\sigma}X^M(\sigma)
e^{ik_N \sqrt{\pi}X^N(\sigma)}
:|B(F)\rangle\nonumber\\
&=&{\cal N}\zeta_{\mu}{-1\over \sqrt{2}}\sum_{n=1}^{\infty}\sqrt{n}\biggl[
{1+{\cal O}^T\over 2}a'{}^{(+)\dagger}_ne^{-in\sigma}-{1+{\cal O}\over 2}a'{}^{(-)\dagger}_ne^{in\sigma}\biggr]^{\mu}\nonumber\\
&&\times \exp\left(k_{\mu}\sum_{n=1}^{\infty}{1\over \sqrt{2n}}\left(
{1+{\cal O}^T\over 2}a'{}^{(+)\dagger}_{n}e^{-in\sigma}
+{1+{\cal O}\over 2}a'{}^{(-)\dagger}_{n}e^{in\sigma}
\right)^{\mu}\right)e^{ik_M\hat{x}^M}|B(F)\rangle\nonumber\\
&=&-\zeta_{\mu}\sum_{n=1}^{\infty}\sqrt{n}\biggl[
{1+{\cal O}^T\over 2}a^{(+)\dagger}_ne^{-in\sigma}-{1+{\cal O}\over 2}a^{(-)\dagger}_ne^{in\sigma}\biggr]^{\mu} 
\label{eq:V_VB}\\
&&\times \exp\left(k_{\mu}\sum_{n=1}^{\infty}{1\over \sqrt{n}}\left(
{1+{\cal O}^T\over 2}a^{(+)\dagger}_{n}e^{-in\sigma}
+{1+{\cal O}\over 2}a^{(-)\dagger}_{n}e^{in\sigma}
\right)^{\mu}\right)e^{ik_M\hat{x}^M}|B(F)\rangle\,. 
\nonumber
\end{eqnarray}
We used $\partial_{\sigma}X$ above instead of
$\partial_{\tau}X$ because we consider ``open string vertex''
in terms of a closed string.

We note that there are no excitations along
Dirichlet directions on $|B(F)\rangle$ in Eqs.(\ref{eq:V_TB}),(\ref{eq:V_VB}).

\section{Computation of vector type fluctuation
\label{sec:vectordetail}
}

Here we present details of computations in \S \ref{sec:vector}.

We first evaluate the quantity ${\cal D}_1$ (\ref{eq:D1_1})
using Eqs.(\ref{eq:sum_cos}-\ref{eq:cosD}).
For the $d^{\sigma+\theta_1}(\cdots)a^{\dagger}$ term in Eq.\eqref{eq:lambdaa},
we have
\begin{eqnarray}
 &&d^{\sigma+\theta_1}\left[{\cal P}_+A^{(1)T}{\cal P}_+
+{\cal P}_-D^{(1)}{\cal P}_-
\right]a^{\dagger}\nonumber\\
&&=-\beta d^{-\beta(\sigma+\theta_1)+\pi}a^{\dagger}\nonumber\\
&&~~~+\left(\sum_{n=1}^{\infty}(-1)^n\cos n(\sigma+\theta_1)+{1\over 2}\right)
{\alpha_1\alpha_2\over \alpha_3}B^TC
\zeta_{\mu} \left({1+{\cal O}^T\over 2}a^{(+)\dagger}
-{1+{\cal O}\over 2}a^{(-)\dagger}\right)^{\mu}\,.
\end{eqnarray}
For the $d^{\sigma+\theta_1}(\cdots){\bf P}$ term, we replace
$a^{\dagger}$ appropriately :
\begin{eqnarray}
  &&d^{\sigma+\theta_1}\left[{\cal P}_+A^{(1)T}{\cal P}_+
+{\cal P}_-D^{(1)}{\cal P}_-
\right]{\alpha_2\over 4}B
\left(
\begin{array}[tb]{c}
 1\\
1
\end{array}
\right)k
\nonumber\\
&&=-i\zeta_{\mu}G^{\mu\nu}k_{\nu}\sum_{m=1}^{\infty}
{\sin m\beta(\sigma+\theta_1)\, \sin m\pi\beta\over \pi m}\\
&&~~~-\zeta_{\mu}\left({\cal O}^T-{\cal O}\over 2\right)^{\mu\nu}k_{\nu}
\left[
\sum_{m=1}^{\infty}{\cos m\beta(\sigma+\theta_1)
\,\sin m\pi\beta\over 2\pi m}
+{1\over 2}(\beta+1)
\left(\sum_{n=1}^{\infty}(-1)^n\cos n(\sigma+\theta_1)+{1\over 2}\right)
\right]\,.\nonumber
\end{eqnarray}
{}In Eq.(\ref{eq:lambda2}), we can compute 
$d^{\sigma+\theta_1}(\cdots)\lambda^{\sigma+\theta_1}$
term using
\begin{eqnarray}
 &&(\sin n\sigma)C^{1\over 2}(1-A^{(1)T}A^{(1)})C^{-{1\over 2}}(\cos n\sigma)
=\sum_{m=1}^{\infty}{\sin 2m\sigma\over 2}
+\sum_{p=1}^{\infty}{\sin\pi p\beta
\sin p\beta\sigma\over \pi p}-\beta\sum_{p=1}^{\infty}
{\sin 2 p\beta\sigma\over 2}\,,~~~~~\\
&& (\cos n\sigma)C^{1\over 2}\left(1
-D^{(1)}D^{(1)T}\right)
C^{-{1\over 2}}(\sin n\sigma)\nonumber\\
&&=\sum_{m=1}^{\infty}{\sin 2 m\sigma\over 2}
+{2\over \pi}\left(\sum_{m=1}^{\infty}(-1)^m\cos m\sigma+{1\over 2}\right)
\sum_{p=1}^{\infty}
{\sin p\pi\beta\,\sin p\beta\sigma\over p}
-\beta\sum_{p=1}^{\infty}{\sin 2p\beta\sigma\over 2}\,.
\end{eqnarray}
Then Eq.(\ref{eq:D1_1}) becomes
\begin{eqnarray}
  &&\mathcal{D}_1=-\beta d^{-\beta(\sigma+\theta_1)+\pi}a^{\dagger}
+\left(\sum_{n=1}^{\infty}\cos n(\pi-\sigma-\theta_1)+{1\over 2}\right)
{\alpha_1\alpha_2\over \alpha_3}B^TC
\zeta_{\mu} \left({1+{\cal O}^T\over 2}a^{(+)\dagger}
-{1+{\cal O}\over 2}a^{(-)\dagger}\right)^{\mu}\nonumber\\
&&-i\zeta_{\mu}G^{\mu\nu}k_{\nu}\biggl[
{2\over \pi}\left(\sum_{m=1}^{\infty}\cos m(\pi-\sigma-\theta_1)+
{1\over 2}\right)\sum_{p=1}^{\infty}
{\sin p\pi\beta\,\sin p\beta(\sigma+\theta_1)\over p}
\biggr]\nonumber\\
&&-\zeta_{\mu}\left({\cal O}^T-{\cal O}\over 2\right)^{\mu\nu}k_{\nu}
\left[
\sum_{m=1}^{\infty}{\cos m\beta(\sigma+\theta_1)
\,\sin m\pi\beta\over 2\pi m}
+{1\over 2}(\beta+1)
\left(\sum_{n=1}^{\infty}\cos n(\pi-\sigma-\theta_1)+{1\over 2}\right)
\right]\nonumber\\
&&=-\beta d^{-\beta(\sigma+\theta_1)+\pi}a^{\dagger}
+\delta(\pi-\sigma-\theta_1)\pi
{\alpha_1\alpha_2\over \alpha_3}B^TC
\zeta_{\mu} \left({1+{\cal O}^T\over 2}a^{(+)\dagger}
-{1+{\cal O}\over 2}a^{(-)\dagger}\right)^{\mu} \label{eq:D1}\\
&&-i\zeta_{\mu}G^{\mu\nu}k_{\nu}\biggl[
2\delta(\pi-\sigma-\theta_1)
\sum_{p=1}^{\infty}
{\sin^2 p\pi\beta\over p}
\biggr]
-\zeta_{\mu}\left({\cal O}^T-{\cal O}\over 2\right)^{\mu\nu}k_{\nu}
\left[
-{\beta+1\over 4}
+{\pi\over 2}(\beta+1)\delta(\pi-\sigma-\theta_1)
\right]\,,\nonumber\\
&&~~~~~~~~~~~~~~~~~~~~~~~~~~~~~~~~~~
(-\pi<\sigma+\theta_1\le \pi)\nonumber
\end{eqnarray}
where we used formulae
\begin{eqnarray}
&&{1\over 2\pi}+ {1\over \pi}\sum_{n=1}^{\infty}\cos nx
=\sum_{n=-\infty}^{\infty}\delta(x-2n\pi)\,,~~~
\sum_{n=1}^{\infty}{\sin nx \cos ny\over n}=\left\{
\begin{array}[tb]{cc}
 -x/2&~~[0\le x < y] \\
(\pi-x)/2&~~[y < x \le  \pi]
\end{array}
\right.\,.~~~~~~
\end{eqnarray}
Similarly, we can evaluate $\mathcal{D}_2$ as
\begin{eqnarray}
\mathcal{D}_2&=&(\beta+1) d^{-(\beta+1)(\pi-\sigma-\theta_2)}a^{\dagger}
-\delta(\sigma+\theta_2)\pi
{\alpha_1\alpha_2\over \alpha_3}B^TC
\zeta_{\mu} \left({1+{\cal O}^T\over 2}a^{(+)\dagger}
-{1+{\cal O}\over 2}a^{(-)\dagger}\right)^{\mu}\nonumber\\
&&+i\zeta_{\mu}G^{\mu\nu}k_{\nu}\biggl[
2\delta(\sigma+\theta_2)
\sum_{p=1}^{\infty}
{\sin^2 p\beta\pi\over p}
\biggr]
+\zeta_{\mu}\left({\cal O}^T-{\cal O}\over 2\right)^{\mu\nu}k_{\nu}
\biggl[
-{\beta\over 4}+
{\pi\over 2}\beta\delta(\sigma+\theta_2)
\biggr]
\,.\label{eq:D2}\\
&&~~~~~~~~~~~~~~~~~~~~~~~~~~~~~~~~~
(-\pi<\pi-\sigma-\theta_2\le \pi)\nonumber
\end{eqnarray}
{}From Eqs.(\ref{eq:VB*B})(\ref{eq:B*VB})(\ref{eq:D1})(\ref{eq:D2}),
we have obtained
\begin{eqnarray}
&&|(\delta_V \Phi_B(\alpha_1))*\Phi_B(\alpha_2)\rangle 
+ |\Phi_B(\alpha_1)*(\delta_V \Phi_B(\alpha_2))\rangle
\nonumber\\
&=&
((-\beta)^{{1\over 2}k_{\mu}G^{\mu\nu}k_{\nu}+1}
+(1+\beta)^{{1\over 2}k_{\mu}G^{\mu\nu}k_{\nu}+1})
c_B|\delta_V\Phi_B(\alpha_1+\alpha_2)\rangle\nonumber\\
&&+{1\over 4}\zeta_{\mu}
\left({{\cal O}^T-{\cal O}\over 2}\right)^{\mu\nu}k_{\nu}
((1+\beta)(-\beta)^{{1\over 2}k_{\mu}G^{\mu\nu}k_{\nu}}
-\beta(1+\beta)^{{1\over 2}k_{\mu}G^{\mu\nu}k_{\nu}})
c_B
|\delta_T\Phi_B(\alpha_1+\alpha_2)\rangle\nonumber\\
&&+
(-\beta)^{{1\over 2}k_{\mu}G^{\mu\nu}k_{\nu}}
c_B\wp\int_{-\pi}^{\pi}{d\sigma'\over 2\pi}
\delta(\pi-\sigma')\biggl[
\pi{\alpha_1\alpha_2\over \alpha_3}\zeta B^TC
\left({1+{\cal O}^T\over 2}a^{(+)\dagger}-{1+{\cal O}\over 2}
a^{(-)\dagger}\right)\nonumber\\
&&-2i\zeta_{\mu}G^{\mu\nu}k_{\nu}\sum_{p=1}^{\infty}{\sin^2 p\pi \beta\over p}
-\zeta_{\mu}
\left({{\cal O}^T-{\cal O}\over 2}\right)^{\mu\nu}k_{\nu}{\pi\over 2}
(\beta+1)\biggr]
e^{-\lambda^{-\beta\sigma'+\pi}a^{\dagger}}|\Phi_B(\alpha_1+\alpha_2)
\rangle\nonumber\\
&&-
(1+\beta)^{{1\over 2}k_{\mu}G^{\mu\nu}k_{\nu}}
c_B\wp\int_0^{2\pi}{d\sigma'\over 2\pi}
\delta(\sigma')\biggl[\pi{\alpha_1\alpha_2\over \alpha_3}\zeta B^TC
\left({1+{\cal O}^T\over 2}a^{(+)\dagger}-{1+{\cal O}\over 2}a^{(-)\dagger}
\right)\nonumber\\
&&-2i\zeta_{\mu}G^{\mu\nu} k_{\nu}\sum_{p=1}^{\infty}{\sin^2 p\pi \beta\over p}
-\zeta_{\mu}
\left({{\cal O}^T-{\cal O}\over 2}\right)^{\mu\nu}k_{\nu}{\pi\over 2}\beta
\biggr]
e^{-\lambda^{-(\beta+1)(\pi-\sigma')}a^{\dagger}}|\Phi_B(\alpha_1+\alpha_2)
\rangle
\nonumber\\
&=&
((-\beta)^{{1\over 2}k_{\mu}G^{\mu\nu}k_{\nu}+1}
+(1+\beta)^{{1\over 2}k_{\mu}G^{\mu\nu}k_{\nu}+1})
c_B|\delta_V\Phi_B(\alpha_1+\alpha_2)\rangle\nonumber\\
&&+{1\over 4}\zeta_{\mu}
\left({{\cal O}^T-{\cal O}\over 2}\right)^{\mu\nu}k_{\nu}
((1+\beta)(-\beta)^{{1\over 2}k_{\mu}G^{\mu\nu}k_{\nu}}
-\beta(1+\beta)^{{1\over 2}k_{\mu}G^{\mu\nu}k_{\nu}})
c_B
|\delta_T\Phi_B(\alpha_1+\alpha_2)\rangle\nonumber\\
&&+
(-\beta)^{{1\over 2}k_{\mu}G^{\mu\nu}k_{\nu}}c_B\wp
\biggl[
{1\over 2}\pi{\alpha_1\alpha_2\over \alpha_3}\zeta B^TC
\left({1+{\cal O}^T\over 2}a^{(+)\dagger}-{1+{\cal O}\over 2}
a^{(-)\dagger}\right)\nonumber\\
&&-i\zeta_{\mu}G^{\mu\nu}k_{\nu}\sum_{p=1}^{\infty}
{\sin^2 p\pi \beta\over \pi p}
-\zeta_{\mu}
\left({{\cal O}^T-{\cal O}\over 2}\right)^{\mu\nu}k_{\nu}{1\over 4}
(\beta+1)\biggr]
e^{-\lambda^{-(\beta+1)\pi+2\pi}a^{\dagger}}|\Phi_B(\alpha_1+\alpha_2)
\rangle\nonumber\\
&&-
(1+\beta)^{{1\over 2}k_{\mu}G^{\mu\nu}k_{\nu}}
c_B\wp
\biggl[{1\over 2}{\alpha_1\alpha_2\over \alpha_3}\zeta B^TC
\left({1+{\cal O}^T\over 2}a^{(+)\dagger}-{1+{\cal O}\over 2}a^{(-)\dagger}
\right)\nonumber\\
&&-i\zeta_{\mu}G^{\mu\nu} k_{\nu}\sum_{p=1}^{\infty}
{\sin^2 p\pi \beta\over \pi p}
-\zeta_{\mu}
\left({{\cal O}^T-{\cal O}\over 2}\right)^{\mu\nu}k_{\nu}{1\over 4}\beta
\biggr]
e^{-\lambda^{-(\beta+1)\pi}a^{\dagger}}|\Phi_B(\alpha_1+\alpha_2)\rangle
\label{eq:vector_LHS}
\nonumber\\
&=&
((-\beta)^{{1\over 2}k_{\mu}G^{\mu\nu}k_{\nu}+1}
+(1+\beta)^{{1\over 2}k_{\mu}G^{\mu\nu}k_{\nu}+1})
c_B|\delta_V\Phi_B(\alpha_1+\alpha_2)\rangle\nonumber\\
&&+((-\beta)^{{1\over 2}k_{\mu}G^{\mu\nu}k_{\nu}}
-(1+\beta)^{{1\over 2}k_{\mu}G^{\mu\nu}k_{\nu}}
)c_B\wp\biggl[{1\over 2}{\alpha_1\alpha_2\over \alpha_3}\zeta B^TC
\left({1+{\cal O}^T\over 2}a^{(+)\dagger}-{1+{\cal O}\over 2}a^{(-)\dagger}
\right)\nonumber\\
&&-i\zeta_{\mu}G^{\mu\nu} k_{\nu}\sum_{p=1}^{\infty}
{\sin^2 p\pi \beta\over \pi p}
\biggr]
e^{-\lambda^{-(\beta+1)\pi}a^{\dagger}}|\Phi_B(\alpha_1+\alpha_2)\rangle
\,.
\end{eqnarray}
Here we adjusted the $2\pi$ interval of integration to validate
summation formulae which we used in computations.


\end{document}